\documentclass[prb,twocolumn,showpacs]{revtex4}
\usepackage{color,soul}
\usepackage{amsmath}
\usepackage{dcolumn}
\usepackage{graphicx}
\usepackage{bm}
\usepackage{amssymb}
\usepackage{subfigure}
\begin{document}

\title{Theory of the anomalous spin dynamics of spin-$\frac{1}{2}$ triangular lattice Heisenberg antiferromagnet and its application to Ba$_3$CoSb$_2$O$_9$}
\author{Chun Zhang and Tao Li}
\affiliation{Department of Physics, Renmin University of China, Beijing 100872, P.R.China}

\begin{abstract}
Although it is well accepted that the  spin-$\frac{1}{2}$ triangular lattice Heisenberg antiferromagnet(TLHAF) has a long range ordered ground state, a thorough understanding of its spin dynamics is still missing. While the linear spin wave theory(LSWT) predicts three branches of magnon mode in the magnetic Brillouin zone(MBZ), the 1/S expansion at the next order is found to break down in a large portion of the MBZ centered around the M point, leaving the fate of the magnon modes there undecided. Recent neutron scattering measurement on Ba$_3$CoSb$_2$O$_9$, an ideal realization of the spin-$\frac{1}{2}$ TLHAF, provides a surprising answer to this issue. Two, rather than three branches of magnon mode are observed around the M point, whose dispersion are strongly renormalized with respect to the LSWT prediction and exhibit pronounced roton-like minimum at the M point. This is accompanied by a strong spin fluctuation continuum at higher energy, inside which two strong and broad spectral peaks of unknown origin are observed. In this work, we propose a simple picture for these spectral anomalies by invoking the resonating valence bond(RVB) physics in the description of the ground state of the system. We find that the roton-like minimum in the magnon dispersion can be explained by the coupling between the collective spin fluctuation and the continuum of Dirac spinon excitation moving in a $\pi$-flux background. We also propose that the two broad peaks in the continuum can be understood respectively as the Landau damped third magnon mode and the Landau damped longitudinal mode. Such a picture can be verified by studying the polarization character of the various spectral features. 
\end{abstract}

\maketitle

\section{Introduction}
Fractionalized spin excitation, or more specifically, spinon, is the key concept invented to describe the dynamical properties of quantum spin liquid system\cite{Balents}. It is until very recently that people realize that it may also play an important role in the spin dynamics of a quantum magnet with a long range ordered ground state. For example, in recent inelastic neutron scattering(INS) measurement on the spin-$\frac{1}{2}$ square lattice Heisenberg antiferromagnet(SLHAF) Cu(DCO$_2$)$_2$$\cdot$4D$_2$O (CFTD)\cite{Piazza}, a roton-like minimum in the magnon dispersion is found around $(\pi,0)$, which is accompanied by a spin rotational invariant spectral continuum at higher energy. The observation of such a spin rotational invariant continuum is very unusual from the perspective of the semiclassical linear spin wave theory(LSWT), but may be consistent with a picture involving spinon excitations.

The spin-$\frac{1}{2}$ antiferromagnetic Heisenberg model defined on the triangular lattice(TLHAF) is the first model system proposed for quantum spin liquid physics\cite{Anderson}. Although it is now well accepted that the model has a long ranged ordered ground state, a thorough understanding of its spin dynamics is still missing. Different from quantum magnet with a collinear magnetic order(such as the SLHAF), the spin rotational symmetry is fully broken in the ground state of the TLHAF and a cubic term is allowed by symmetry in the spin wave Hamiltonian. As a result, while LSWT predicts that there should be three branches of magnon mode in the magnetic Brillouin zone(MBZ)\cite{Jolicoeur}, the 1/S correction to the magnon self-energy is found to diverge in a large portion of the magnetic Brillouin zone centered around the M point\cite{Zhitomirsky}.  Much efforts has be devoted to go beyond the understanding provided by LSWT but the general picture about the fate of the magnon around the M point is still elusive\cite{Mourigal,Ghioldi,Ghioldi1,Pollmann,Becca}. In particular, it is not clear if coherent magnon mode can survive in this momentum region and how is their dispersion related to the LSWT predictions. 

Recently, inelastic neutron scattering(INS) measurement on a prototype of spin-$\frac{1}{2}$ TLHAF material, Ba$_3$CoSb$_2$O$_9$, provides a surprising answer to the above problem\cite{Shirata,Susuki,Zhou,Ma,Saya,Kamiya}. In particular, it is found in [\onlinecite{Saya}] in that there are two, rather than three branches of coherent magnon mode around the M point, whose dispersion are significantly softened as compared to the prediction of the LSWT and exhibit pronounced roton-like minimum at the M point.  Similar to the case of SLHAF, such roton-like minimum is accompanied by strong spin fluctuation continuum at higher energy. In addition, there are two broad spectral peaks of unknown origin in this continuum, which seem to be connected to two dispersive modes outside the most anomalous region around the M point. 

The observation of these spectral anomalies raises the following questions. First,  what is the origin of the roton-like minimum and the accompanying spin fluctuation continuum around the M point? Second, what is the relation between the magnon mode observed in the experiment and that predicted by the LSWT ? In particular, where is the third magnon mode predicted by the LSWT?  Third, what is the origin of the two broad spectral peaks in the spin fluctuation continuum? What is the meaning of their dispersion? Finally, is there any simple explanation why the observed spectral anomalies is so strongly momentum dependent and significant only around the M point?

The purpose of this work is to propose a simple understanding of these observations by invoking resonating valence bond(RVB) physics in the description of the ground state of the system. Besides the semiclassical 120 degree ordering pattern, we assume the existence in the ground state of the spin-$\frac{1}{2}$ TLHAF a specific RVB structure. Spinon moving on such an RVB background will experience a gauge flux of $\pi$ around each elementary plaquette of the triangular lattice and will thus be endowed a Dirac-type dispersion. The most important observation of this work is that the momentum at the M point, where the spin dynamics of the spin-$\frac{1}{2}$ TLHAF is the most anomalous, is just the momentum difference of the two Dirac nodes of the spinon dispersion in such a $\pi$-flux phase.

As we will show in this paper, the dramatic change in the spin dynamics of the spin-$\frac{1}{2}$ TLHAF around the M point can be attributed to the particle-hole excitation of the spinon system between its two Dirac nodes. In particular, we find that the roton-like minimum of the magnon dispersion around the M point observed in Ba$_3$CoSb$_2$O$_9$ can be attributed to the level repulsion effect between the spinon continuum and the collective spin excitation below it. We also propose that the two broad spectral peaks in the continuum can be understood as remanant of the third magnon mode and the longitudinal spin fluctuation mode swallowed by the spinon continuum around the M point. We argue that the anomaly in the high energy spin fluctuation spectrum of a frustrated quantum magnet can serve as a useful diagonose of the RVB structure in its quantum ground state.

Unlike quantum magnet with a collinear ordering pattern, the transverse and longitudinal spin fluctuation are intrinsically entangled with each other in the spin fluctuation spectrum of the spin-$\frac{1}{2}$ TLHAF. The spin fluctuation in a quantum magnet with a non-collinear ordering pattern can thus have a complicated polarization character that depends on both momentum and energy. Such polarization information can be crucial for identifying the nature of various spectral features observed in the experiment. A detailed prediction for the polarization character of the various modes of the spin-$\frac{1}{2}$ TLHAF is given in our work, which can be verified in future polarized INS measurement. In particular, we predict that the highest dispersive mode in the spectrum is a longitudinal spin fluctuation mode.  

The paper is organized as follows. In the next section, we introduce a Fermionic spinon theory for the ground state and the spin dynamics of the spin-$\frac{1}{2}$ TLHAF.  In the third section, we present the spin fluctuation spectrum calculated from the spinon theory and compare it with the INS results on Ba$_3$CoSb$_2$O$_9$. In the last section, we discuss the implications of our results and propose possible experimental verification of the theory.

\section{A Fermionic RVB theory of the spin-$\frac{1}{2}$ TLHAF}
The spin-$\frac{1}{2}$ TLHAF reads
\begin{equation}
H=J\sum_{<i,j>}\mathbf{S}_{i}\cdot \mathbf{S}_{j},
\end{equation}
in which the sum is over nearest neighboring sites of the triangular lattice. In the LSWT, the spin excitation is described by a spin-1 Bosonic operator, the quanta of deviation of the local spins from their expectation value in the semiclassical ground state. Anticipating the inadequacy of such a semiclassical description, in this study we represent the spin in terms of Fermionic spinon operator as $\mathbf{S}_{i}=\frac{1}{2}\sum_{\alpha,\beta}f^{\dagger}_{i,\alpha}\bm{\sigma}_{\alpha,\beta}f_{i,\beta}$. Here $\alpha,\beta=\pm1$ is the spin index of the spinon operator, $\bm{\sigma}$ is the usual Pauli matrix. The Fermionic spinon operator should satisfy the constraint of $\sum_{\alpha}f^{\dagger}_{i,\alpha}f_{i,\alpha}=1$ to be a faithful representation of the spin algebra. In the spinon formulation, the Heisenberg Hamiltonian reads
\begin{equation}
H=\frac{J}{4}\sum_{<i,j>,\alpha,\beta}[\ 2f^{\dagger}_{i,\alpha}f_{i,\beta} f^{\dagger}_{j,\beta}f_{j,\alpha}\ -\ n_{i,\alpha}n_{j,\beta}],
\end{equation}
in which $n_{i,\alpha}=f^{\dagger}_{i,\alpha}f_{i,\alpha}$. 

In the spinon representation, we can define two types of mean field order parameter for the ground state of the system. The first type is just the conventional ordered moment $\mathbf{m}_{i}=\frac{1}{2}\sum_{\alpha,\beta}\langle f^{\dagger}_{i,\alpha}\bm{\sigma}_{\alpha,\beta}f_{i,\beta}\rangle$ defined on a single site. The second type is the RVB correlation between a pair of sites. The RVB order parameter can take either the form of a hopping amplitude $\chi_{i,j}=\sum_{\alpha}\langle f^{\dagger}_{i,\alpha}f_{j,\alpha}\rangle$, or that of a pairing amplitude $\Delta_{i,j}=\sum_{\alpha}\langle \alpha f^{\dagger}_{i,\alpha}f^{\dagger}_{j,-\alpha}\rangle$. The structure of such RVB order parameters should be classified by the so called projective symmetry group technique since it involves gauge degree of freedom. In particular, an RVB mean field ansatz that is gauge equivalent to an ansatz with only hopping-type RVB parameter is called a U(1) RVB ansatz\cite{Wen}.

According to previous studies, the most favorite RVB structure for the ground state of the spin-$\frac{1}{2}$ TLHAF is described by a U(1) mean field ansatz with a gauge flux of $\pi$ enclosed in every elementary plaquette of the triangular lattice\cite{Yunoki}. The mean field ansatz reads
\begin{equation}
H_{\mathrm{RVB}}=-\sum_{<i,j>,\alpha}\eta_{i,j}f^{\dagger}_{i,\alpha}f_{j,\alpha}.
\end{equation} 
Here we have set the magnitude of the spinon hopping integral as the unit of energy. $\eta_{i,j}=\pm1$ is the sign of the hopping integral between site $i$ and $j$, chosen in such a way that each elementary plaquette of the triangular lattice encloses a gauge flux of $\pi$. In the presence of the $\pi$-flux, the unit cell of the spinon Hamiltonian is doubled.  For the particular choice of $\eta_{i,j}$ shown in Fig.1a, the spinon dispersion is given by 
\begin{equation}
\epsilon_{\mathbf{k}}=\pm 2\sqrt{\cos^{2}(k_{1})+\cos^{2}(k_{2})+\sin^{2}(k_{1}-k_{2})}.
\end{equation}
Here $\mathbf{k}=k_{1}\mathbf{b}_{1}+k_{2}\mathbf{b}_{2}$, with $(k_{1},k_{2})\in [0,\pi]\otimes[0,2\pi]$. $\mathbf{b}_{1}$ and $\mathbf{b}_{2}$  are the two reciprocal vectors of the triangular lattice. The dispersion is particle-hole symmetric and there are two Dirac nodes in the reduced Brillouin zone at $\mathbf{k}_{1}=(\frac{\pi}{2},\frac{\pi}{2})$ and $\mathbf{k}_{2}=(\frac{\pi}{2},\frac{3\pi}{2})$. The key observation of this study is that the M point in the Brillouin zone, where the spin dynamics of the spin-$\frac{1}{2}$ TLHAF is the most anomalous, is just the difference between $\mathbf{k}_{1}$ and $\mathbf{k}_{2}$.

To describe the collective spin fluctuation in the system, we introduce phenomenologically a Hubbard-type local correlation of the form $H_{U}=U\sum_{i}n_{i,\uparrow}n_{i,\downarrow}$ between the spinons. Such a local correlation takes into account partially the effect of the no double occupancy constraint on the spinon.  This trick has been used successfully in the study of the SLHAF before\cite{Ho}. Here we use it to induce an ordered magnetic moment with the 120 degree ordering pattern. The mean field Hamiltonian of the spinon system in the presence of the static magnetic order reads
\begin{equation}
H_{\mathrm{MF}}=-\sum_{<i,j>,\alpha}\eta_{i,j}f^{\dagger}_{i,\alpha}f_{j,\alpha} -\frac{2U}{3}\sum_{i,\alpha,\beta}\mathbf{m}_{i}\cdot f^{\dagger}_{i,\alpha}\bm{\sigma}_{\alpha,\beta}f_{i,\beta}.
\end{equation} 
The magnitude of the ordered moment can be determined from the self-consistent equation $m=|\mathbf{m}_{i}|=\frac{1}{2} \sum_{\alpha,\beta} \langle \ f^{\dagger}_{i,\alpha}\bm{\sigma}_{\alpha,\beta}f_{i,\beta}\ \rangle_{\mathrm{MF}}$, whose solution is shown in Fig.1b. For $U>U_{c}\approx 9.27$, a finite $m$ appears in the mean field ground state. We note that the unit cell of $H_{\mathrm{MF}}$ is six times larger than the unit cell of the triangular lattice as a result of both the $\pi$-flux structure and the 120 degree order(see Fig.1a).
  
\begin{figure}
\includegraphics[width=8.5cm]{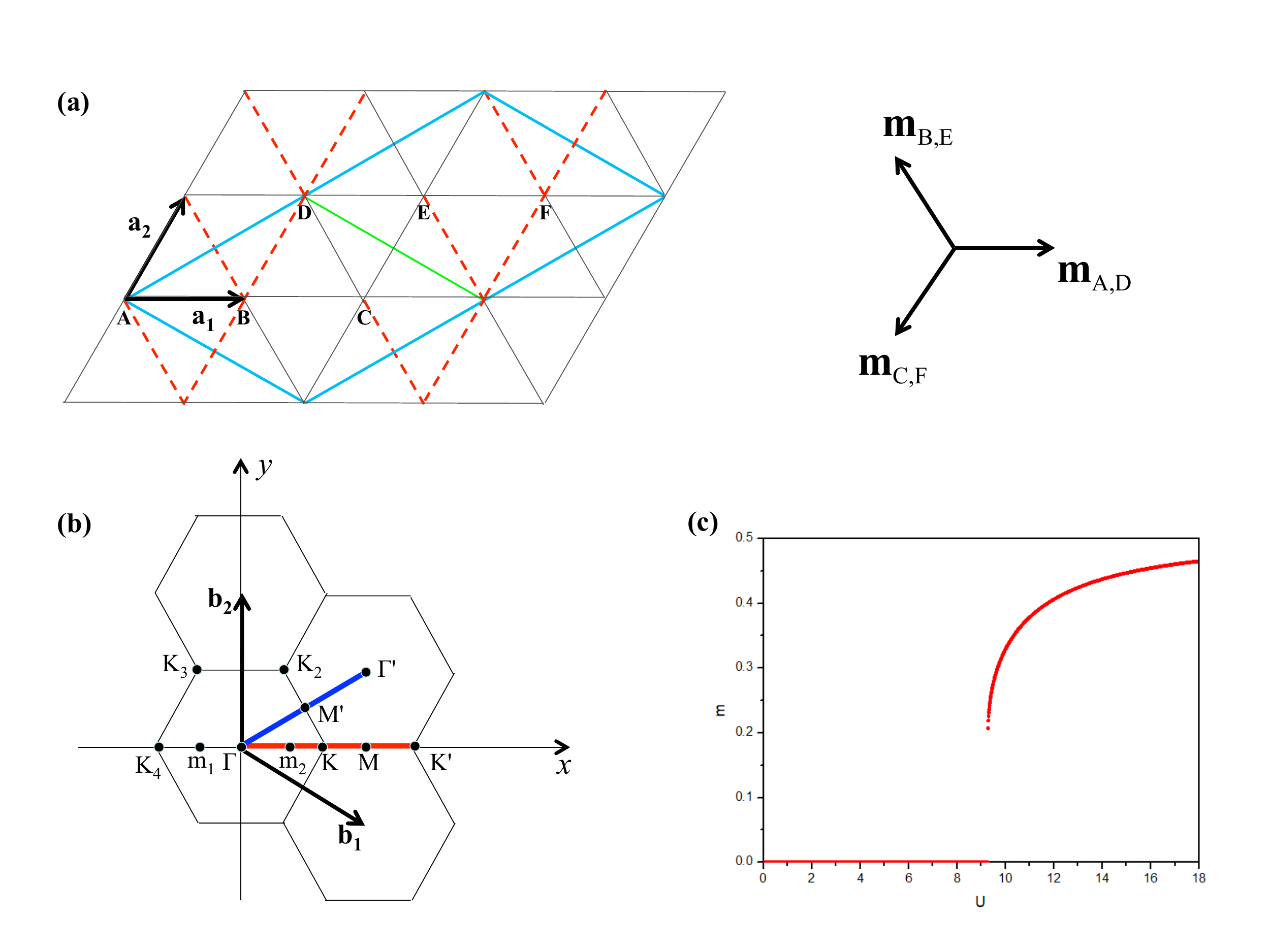}
\caption{Illustration of the spinon model studied in this work. (a)The unit cell of the spinon Hamiltonian is given by the region enclosed by the blue parallelogram, which contains 6 sites of the triangular lattice. The sign of the spinon hopping integral $\eta_{i,j}$ is negative on the red dashed bonds and is otherwise positive, so that a $\pi$-flux is enclosed in each elementary plaquette of the triangular lattice. The direction of the ordered moment on the 6 sites in the spinon unit cell is shown on the right. The magnetic unit cell is the half of the spinon unit cell. (b) The lattice Brillouin zone and the high symmetry points in the momentum space that will be referred to below. We will focus on the spin dynamics of the system on the red path($\Gamma-\mathbf{K}-\mathbf{K'}$) and the blue path($\Gamma-\mathbf{M'}-\Gamma'$) in this study. $\mathbf{b}_{1}$ and $\mathbf{b}_{2}$ are the two reciprocal vectors of the triangular lattice and statisfy $\mathbf{a}_{i}\cdot\mathbf{b}_{j}=\delta_{i,j}$. (c) The self-consistent solution of the mean field equation for the ordered moment $m$  as a function of $U$ in the presence of the $\pi$-flux.}
\end{figure}

 The spin fluctuation spectrum of the system can be extracted from its dynamical spin susceptibility. Here we will adopt a global coordinate system for the spin, in which the dynamical spin susceptibility takes the form of a $18\times18$ matrix
 \begin{equation}
\bm{\chi}_{\mu,\nu}^{i,j}(\mathbf{q},\tau)=-\langle \ T_{\tau} \mathrm{S}^{i}_{\mu}(\mathbf{q},\tau)\  \mathrm{S}^{j}_{\nu}(-\mathbf{q},0) \rangle.
 \end{equation}
Here $\mu,\nu=1,..,6$ denotes the index of the six sublattice A, B,..., F in the spinon unit cell. $i,j=x,y,z$ denotes the spin component index. 
\begin{equation}
\mathrm{S}^{i}_{\mu}(\mathbf{q})=\frac{1}{2}\sum_{\mathbf{k},\alpha,\beta}f^{\dagger}_{\mathbf{k+q},\mu,\alpha}\sigma^{i}_{\alpha,\beta}f_{\mathbf{k},\mu,\beta},
\end{equation}
is the spin density operator in the $i$-th direction and $\mu$-th sublattice. At the mean field level, the dynamical spin susceptibility is given by
\begin{equation}
\bm{\chi}_{\mu,\nu}^{(0)i,j}(\mathbf{q},\tau)=\frac{1}{4}\sum_{\mathbf{k}}\sigma^{i}_{\alpha,\beta}\sigma^{j}_{\gamma,\delta}G^{\delta,\alpha}_{\mu\nu}(\mathbf{k+q},-\tau)G^{\beta,\gamma}_{\mu,\nu}(\mathbf{k},\tau),
\end{equation}
in which $G$ is the spinon Green's function calculated at the mean field level. The dynamical spin susceptibility of the interacting system can be obtained by RPA correction on the bare susceptibility $\chi^{(0)}$, which is given by
\begin{equation}
 \bm{\chi}(\mathbf{q},i\omega_{n})=\frac{\bm{\chi}^{(0)}(\mathbf{q},i\omega_{n})}{\mathbf{I}-\frac{4U}{3}\bm{\chi}^{(0)}(\mathbf{q},i\omega_{n})}.
\end{equation}
We note that this equation should be understood as a matrix equation and the RPA kernel $\frac{4U}{3}$ should be understood as $\frac{4U}{3}\mathbf{I}$, in which $\mathbf{I}$ denotes a $18\times18$ identity matrix.

From the Lehmann representation, we can express the dynamical spin susceptibility in terms of the spectral matrix as follows
\begin{equation}
\bm{\chi}^{i,j}_{\mu,\nu}(\mathbf{q},i\omega_{n})=\frac{1}{2\pi}\int d\omega \ \ \frac{\ \mathbf{R}^{i,j}_{\mu,\nu}(\mathbf{q},\omega)\  \ }{\  i\omega_{n}-\omega\  \ },
\end{equation}
in which the spectral matrix is given by
\begin{eqnarray}
\mathbf{R}^{i,j}_{\mu,\nu}(\mathbf{q},\omega)=\frac{1}{Z}\sum_{n,m}e^{-\beta(E_{n}-E_{m})}\langle n| \mathrm{S}^{i}_{\mu}(\mathbf{q})|m\rangle\times\nonumber\\
\langle m |\mathrm{S}^{j}_{\nu}(-\mathbf{q})|n\rangle\times 2\pi \delta(\hbar \omega -(E_{n}-E_{m})).
\end{eqnarray}
Here $Z=\sum_{n}e^{-\beta E_{n}}$ is the partition function of the system. The spectral matrix $\mathbf{R}$ can be extracted from the retarded spin susceptibility matrix by taking its anti-Hermitian part, or
\begin{equation}
\mathbf{R}^{i,j}_{\mu,\nu}(\mathbf{q},\omega)=\frac{1}{2i}\left[ \bm{\chi}^{i,j}_{\mu,\nu}(\mathbf{q},\omega+i0^{+})-\bm{\chi}^{j,i}_{\nu,\mu}(\mathbf{q},\omega-i0^{+})\right].
\end{equation}
It can be shown that the spectral matrix $\mathbf{R}^{i,j}_{\mu,\nu}(\mathbf{q},\omega)$ is positive definite and Hermitian in its index $(i,\mu)$ and $(j,\nu)$. We can thus interpret its eigenvalues and eigenvectors as the spectral weight and polarization vectors of the spin fluctuation at momentum $\mathbf{q}$ and frequency $\omega$.
The intensity measured by the INS experiment(in the unpolarized scattering mode) is related to the spectral matrix as 
\begin{equation}
I(\mathbf{q},\omega)=\sum_{\mu,\nu,i}e^{i\mathbf{q}\cdot(\bm{\delta}_{\mu}-\bm{\delta}_{\nu})}\mathbf{R}^{i,i}_{\mu,\nu}(\mathbf{q},\omega).
\end{equation}
Here $\bm{\delta}_{\mu}$ denotes the position of the $\mu$-th site in the unit cell. Note that we have neglected in this formula the transverse factor that depends on the actual scattering vector $\mathbf{Q}=\mathbf{q}+\mathbf{G}$ adopted in the INS experiment. 
   
We note that an on-site Hubbard interaction is just a first approximation to the spinon interaction. In the previous discussion, we have decoupled the Heisenberg interaction in the RVB channel only. To represent its residual effect in the spin density channel, we reintroduce a Heisenberg interaction of the form $J_{1}\mathbf{S}_{i}\cdot \mathbf{S}_{j}$ between spinons on nearest neighboring sites. We then decouple the Heisenberg interaction in the spin density channel. As a result, the mean field that spinon experience in the spin density channel becomes $(\frac{4U}{3}+3J_{1})|\mathbf{m}|$. The RPA kernel now becomes $\mathbf{V}(\mathbf{q})=\frac{4U}{3}\mathbf{I}-J(\mathbf{q})$. Here $J(\mathbf{q})$ is the Fourier transform of Heisenberg interaction and is a $18\times18$ matrix. We note that since we have set the spinon hopping integral as the unit of energy of the system, both $U$ and $J_{1}$ should be understood as phenomenological parameters in the spinon description.

\section{Results and Discussions}
In this section, we will present the spin fluctuation spectrum of the spin-$\frac{1}{2}$ TLHAF calculated from the spinon theory detailed in the last section. To have a better understanding on how the spin dynamics of the system is influenced by the spinon excitation, we will first present a detailed discussion on the structure of the bare spin susceptibility of the spinon system in subsection A. This is followed by the results of the RPA-corrected spin fluctuation spectrum in subsection B. The comparison with the INS results on Ba$_{3}$CoSb$_{2}$O$_{9}$ is done in subsection C.

\subsection{Structure of the bare spin susceptibility}
In this study, we will focus on the spin dynamics at zero temperature. The spectral matrix of the bare spin susceptibility of the spinon system is given by
\begin{equation}
\mathbf{R}^{(0),i,j}_{\mu,\nu}(\mathbf{q},\omega)=\frac{\pi}{4}\sum_{\mathbf{k},m,n} M^{i,j,m,n}_{\mu,\nu,\mathbf{k},\mathbf{k'}}\ \delta ( \hbar\omega-\epsilon^{n}_{\mathbf{k'}}+\epsilon^{m}_{\mathbf{k}} ),
\end{equation}
in which $\mathbf{k'}=\mathbf{k-q}$ and $\epsilon^{m}_{\mathbf{k}}<0, \ \epsilon^{n}_{\mathbf{k'}}>0$. The matrix element is given by
\begin{eqnarray}
M^{i,j,m,n}_{\mu,\nu,\mathbf{k},\mathbf{k'}}=\sum_{\alpha,\beta,\gamma,\delta} (u^{m*}_{\mu,\alpha,\mathbf{k}}\sigma^{i}_{\alpha,\beta}u^{n}_{\mu,\beta,\mathbf{k'}})\times(u^{n*}_{\nu,\gamma,\mathbf{k'}}\sigma^{j}_{\gamma,\delta}u^{m}_{\nu,\delta,\mathbf{k}}).\nonumber
\end{eqnarray}
Here $u^{m}_{\mu,\alpha,\mathbf{k}}$ is the eigenvector of the spinon mean field Hamiltonian at momentum $\mathbf{k}$ and with eigenvalue $\epsilon^{m}_{\mathbf{k}}$ . 

The spectral intensity $I(\mathbf{q},\omega)$ of the bare spin susceptibility is plotted in Fig.2 along $\Gamma-\mathbf{K}-\mathbf{K}'$ for both $U$=0 and $U$=9.4. The spectrum is composed of the particle-hole continuum of the spinon excitation. When $U$=0, the spinon continuum reaches zero energy with a Dirac-type dispersion in its lower boundary at both the M point and the $\Gamma$ point. As we mentioned before, M point is the momentum transfer of spinon transition between the two Dirac nodes. On the other hand, the Dirac cone structure around the $\Gamma$ point corresponds to the transition within a Dirac node. It can be seen that the Dirac cone structure around the $\Gamma$ point is suppressed in intensity as compared to that around the M point. In particular, the spectral intensity right at the $\Gamma$ point is exactly zero. Such suppression in spectral intensity can be understood either as a result of the matrix element effect of the spinon excitation within the same Dirac node, or, as a result of the spin rotational symmetry of the spinon Hamiltonian at $U$=0. 

When we turn on a finite ordered moment at $U$=9.4, the following changes appear in the bare spin susceptibility. First, the spinon continuum around the M point and the $\Gamma$ point become both gapped. Second, there emerges additional Dirac cone structure in the spinon continuum at the $\mathbf{m}_{2}$ and $\mathbf{K}$ point(and symmetry-related momentum in the Brillouin zone). Third, the range of the continuum in energy is enlarged and an almost non-dispersive structure emerges at the upper boundary of the spinon continuum. Fourth, the spinon continuum around the $\Gamma$ point is enhanced in intensity. In particular, the spectral intensity right at the $\Gamma$ point is now nonzero.  

To understand these changes in the bare spin susceptibility, we plot in Fig.2c and Fig.2d the spinon density of state for $U$=0 and $U$=9.4. When $m$ is nonzero, as is the case for $U$=9.4, a spinon gap opens at the Dirac point. The band width of the spinon spectrum is also found to be enhanced with the opening of the spinon gap. More specifically, the Van Hove singularity in the spinon density of state is found to split into two peaks when $m$ is nonzero, with the peak at higher energy forming an almost non-dispersive structure. This peak is found to be responsible for the non-dispersive feature in the bare spin susceptibility. At the same time, since the spinon momentum is conserved only up to the reciprocal vectors of the magnetic Brillouin zone when $m$ is nonzero, the Dirac cone structure around the M point will be scattered to the $\mathbf{m}_{1}$ and $\mathbf{m}_{2}$ point. Similarly, the Dirac cone structure around the $\Gamma$ point will be scattered to the $\mathbf{K}$ and $\mathbf{K'}$ point. Finally, since the spin rotational symmetry is fully broken when $m$ is nonzero, the spinon transition within a Dirac node is now allowed. This explains why the Dirac cone structure around the $\Gamma$ point is enhanced.

\begin{figure}
\includegraphics[width=4.2cm,height=3cm]{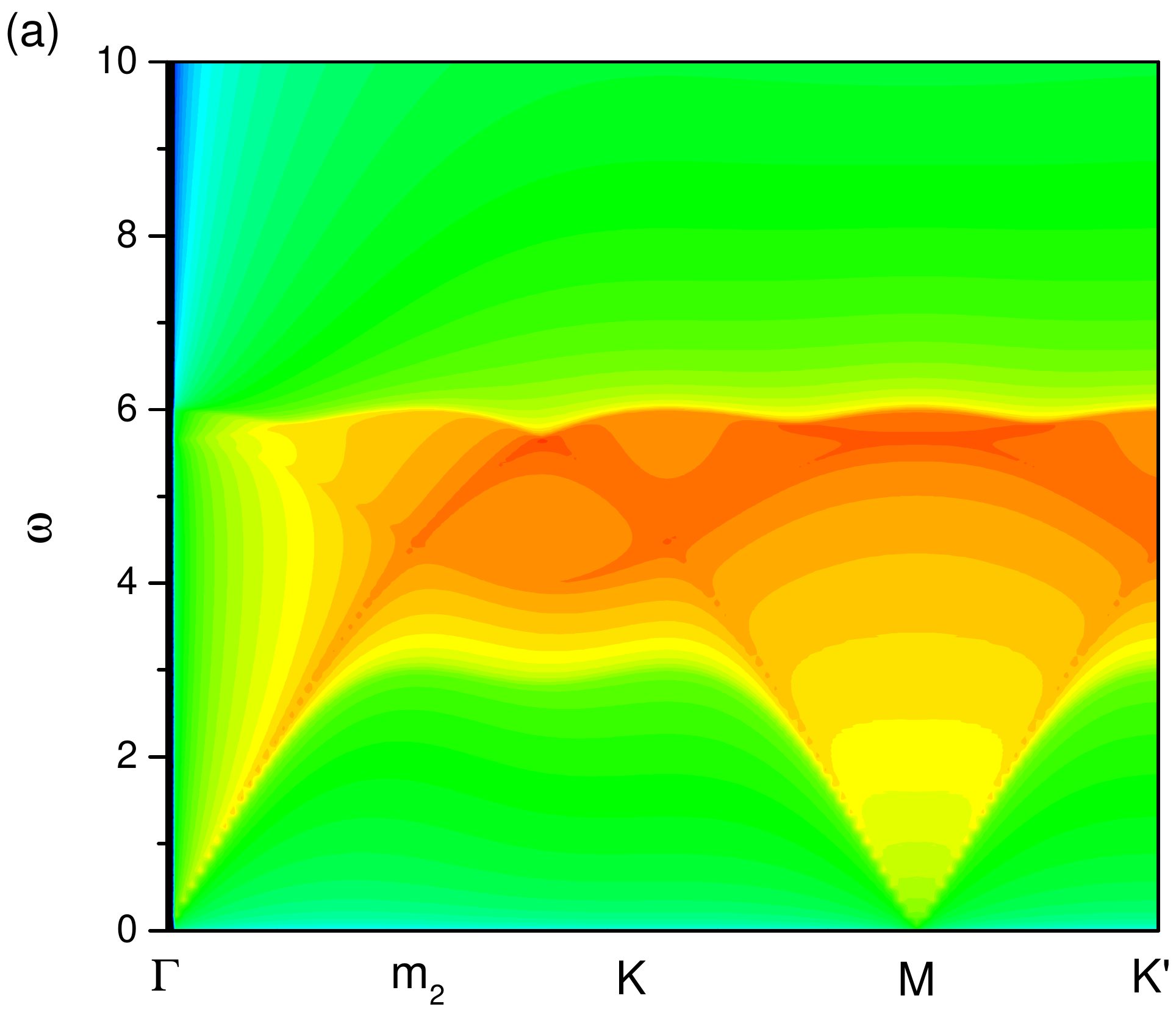}
\includegraphics[width=4.2cm,height=3cm]{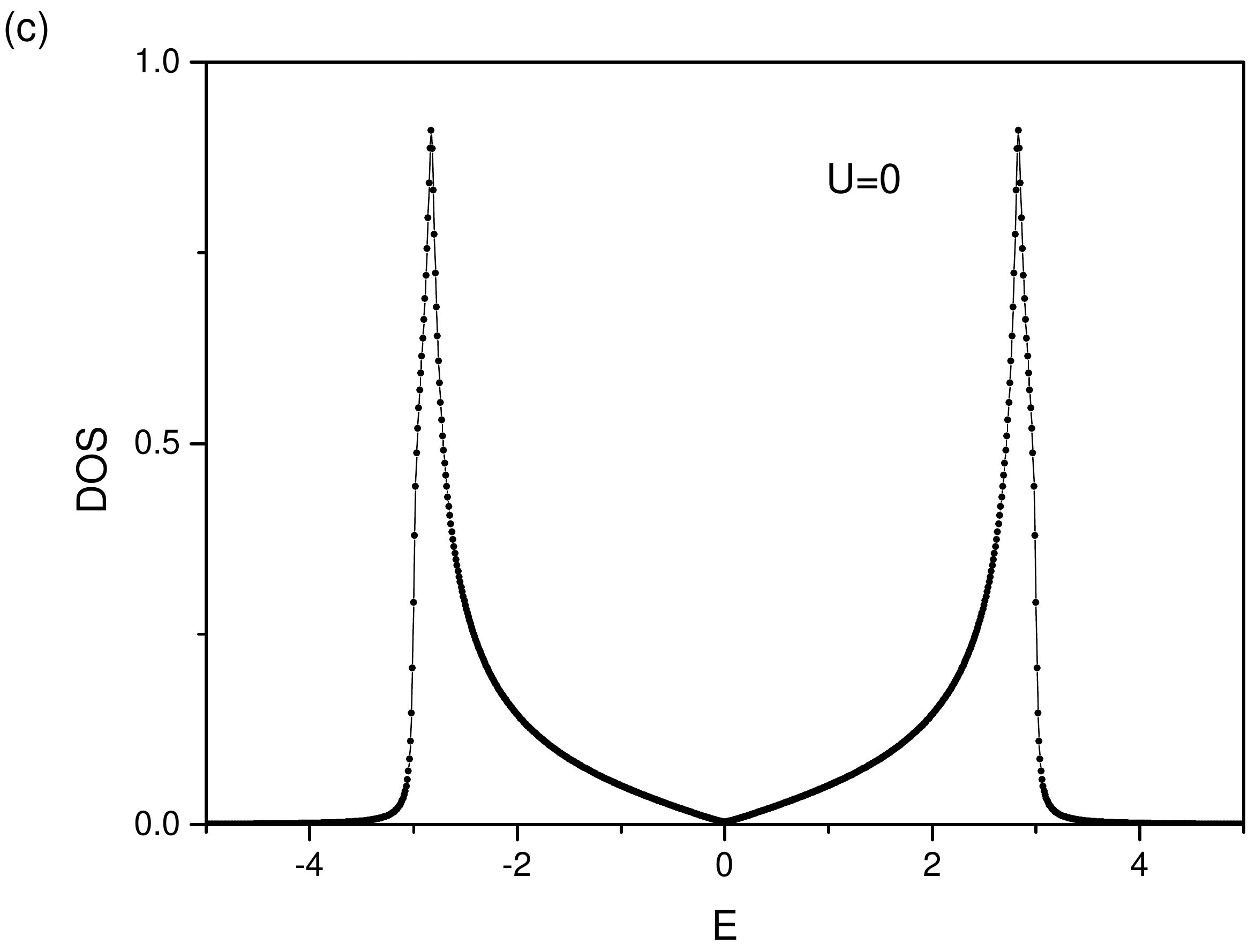}
\includegraphics[width=4.2cm,height=3cm]{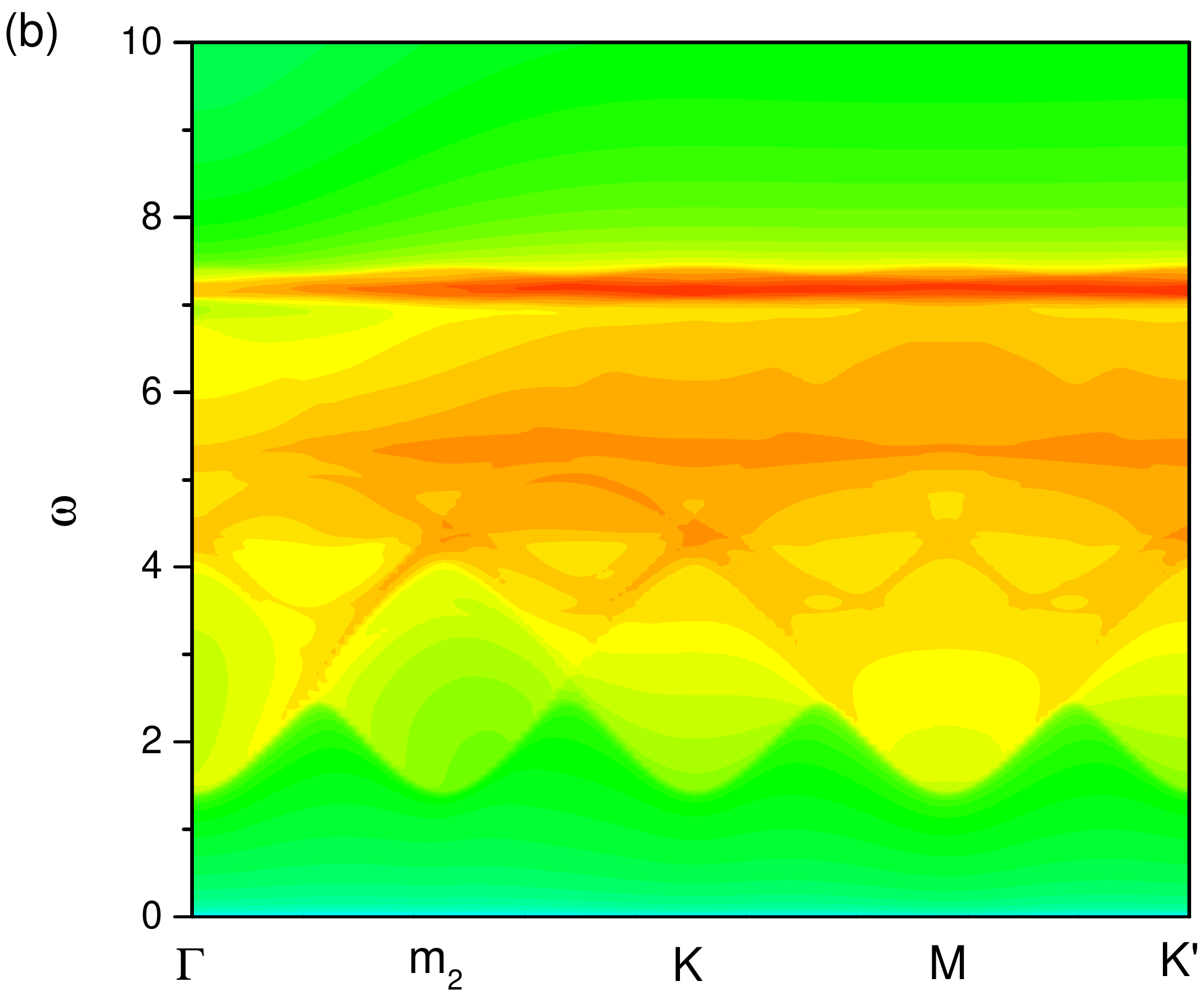}
\includegraphics[width=4.2cm,height=3cm]{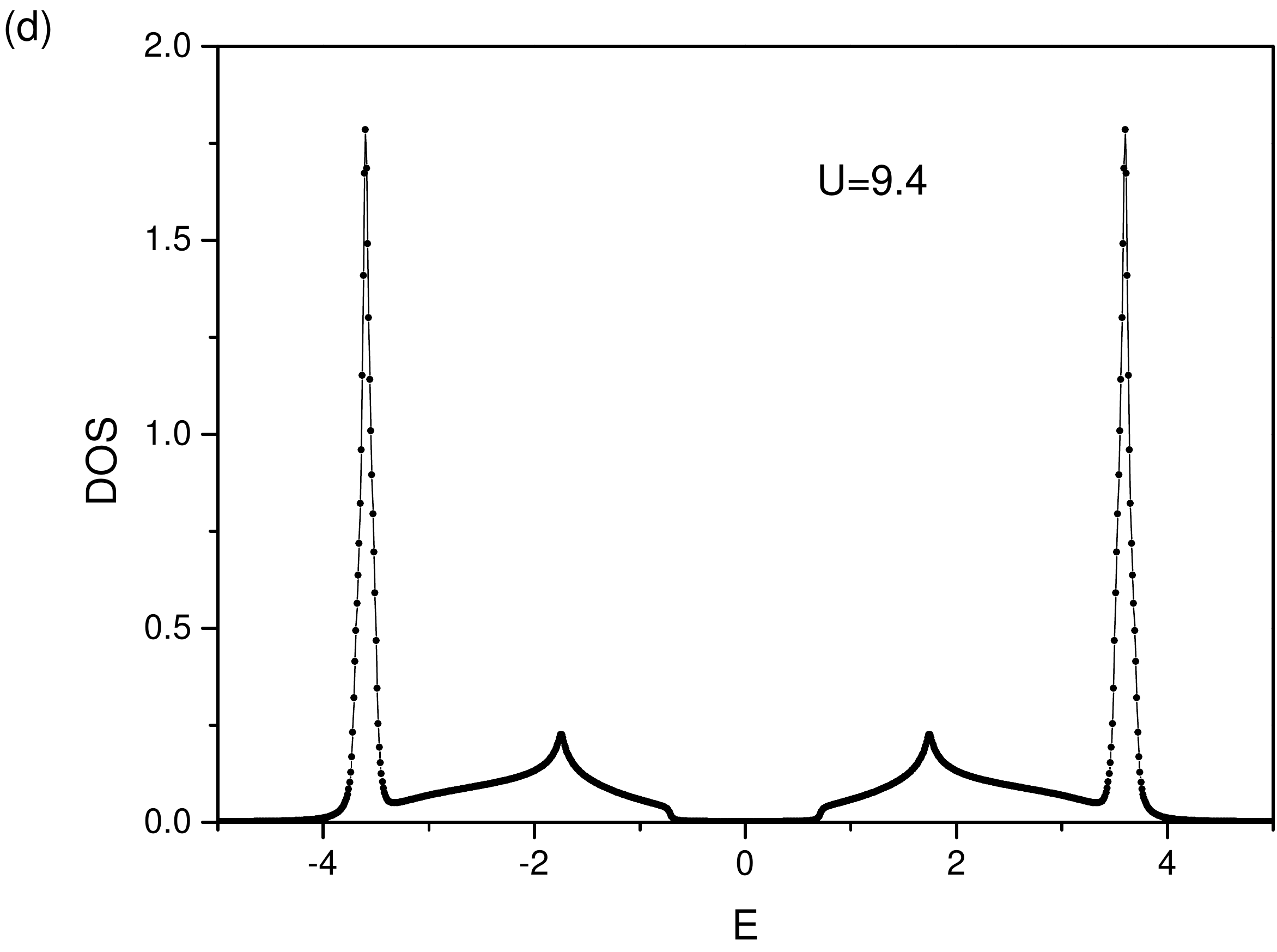}
\caption{Illustration of the bare spectral intensity at $U=0$(a) and $U=9.4$(b). The intensity is plotted in logarithmic scale to highlight the spectral weight around the Dirac cone structures. Plotted on the right are the corresponding spinon density of state(DOS).}
\end{figure}

\subsection{The RPA-corrected spin fluctuation spectrum}
Magnon modes appear below the spinon continuum after we turn on the RPA correction.  We first consider the RPA correction caused by the Hubbard interaction. The evolution of the RPA-corrected spin fluctuation spectrum with $U$ is illustrated in Fig.3 along both $\Gamma-\mathbf{K}-\mathbf{K'}$ and $\Gamma-\mathbf{M}'-\Gamma'$. These spectrums correspond to four values of $U$ above $U_{c}$, namely, $U$=12, $U$=10, $U$=9.4 and $U$=9.3. 

For $U$=12, which is significantly higher than $U_{c}$, the magnon mode lies far below the spinon continuum and exhibits a dispersion that is in close agreement with the prediction of the LSWT. More specifically, there are three branches of magnon in the magnetic Brillouin zone and their energies go correctly to zero at the $\Gamma$ and $\mathbf{K}$ point, as is guaranteed by the self-consistency of the mean field solution. Similar to the prediction of the LSWT, the upper two magnon branches are almost degenerate with each other. A closer inspection indicates that the upper two magnon branches actually cross at the M point. 

With the decrease of $U$ and the lowering of the spinon continuum in energy, new features emerge in the spin fluctuation spectrum. This is best illustrated in Fig3c for the case of $U$=10. First, magnon dispersion starts to deviate from the LSWT prediction, especially around the M point. More specifically, a roton-like minimum starts to develop in the dispersion of the lowest magnon branch at the M point. The dispersion of the upper two magnon modes are also strongly renormalized downward at the M point, resulting in a much more evident splitting between them. However, the degeneracy between the upper two magnon modes at the M point remains intact. Second, a fourth collective mode emerges around the $\mathbf{K}$ point near the bottom of the Dirac cone.  

\begin{figure}
\includegraphics[width=4.2cm,height=3cm]{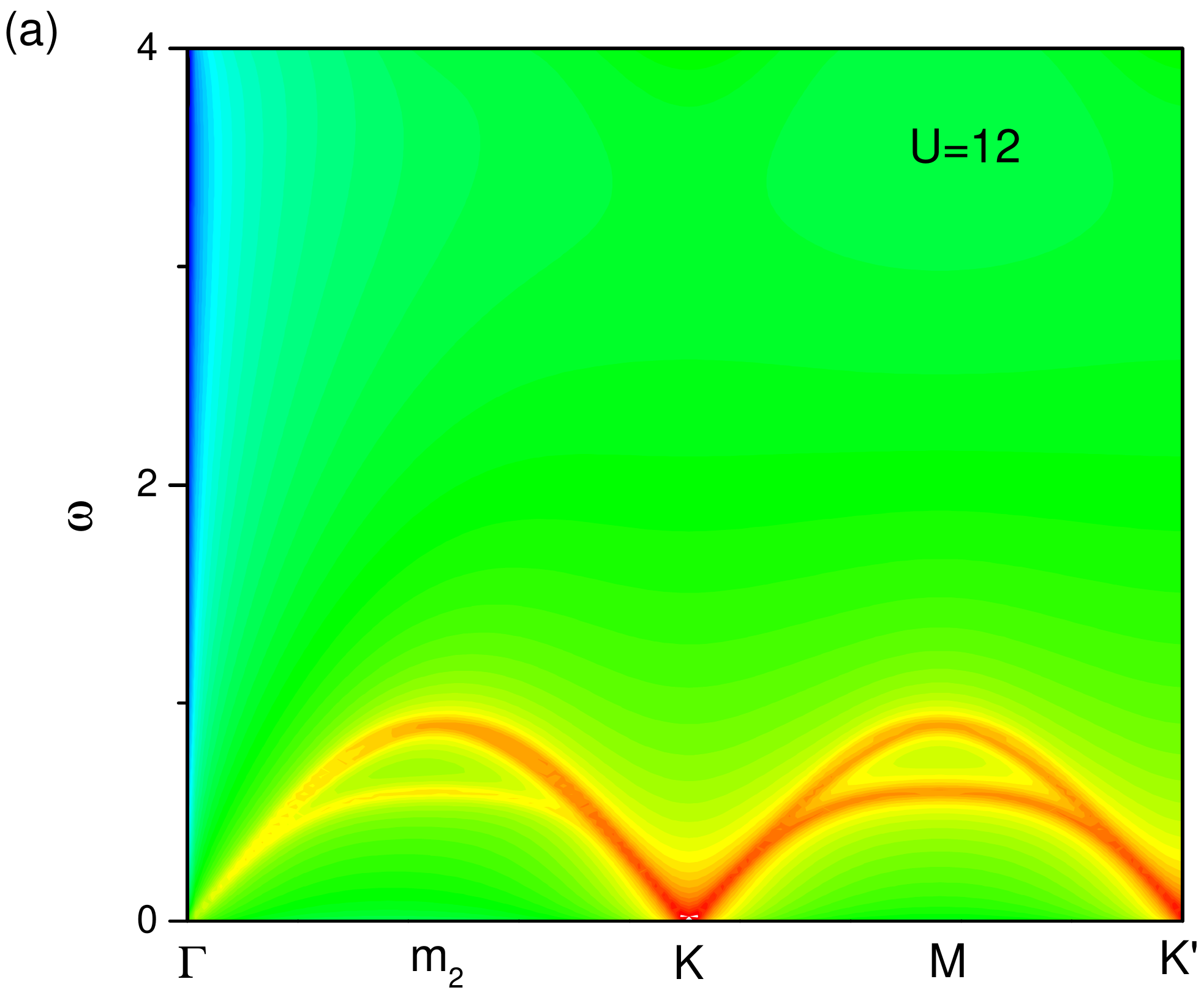}
\includegraphics[width=4.2cm,height=3cm]{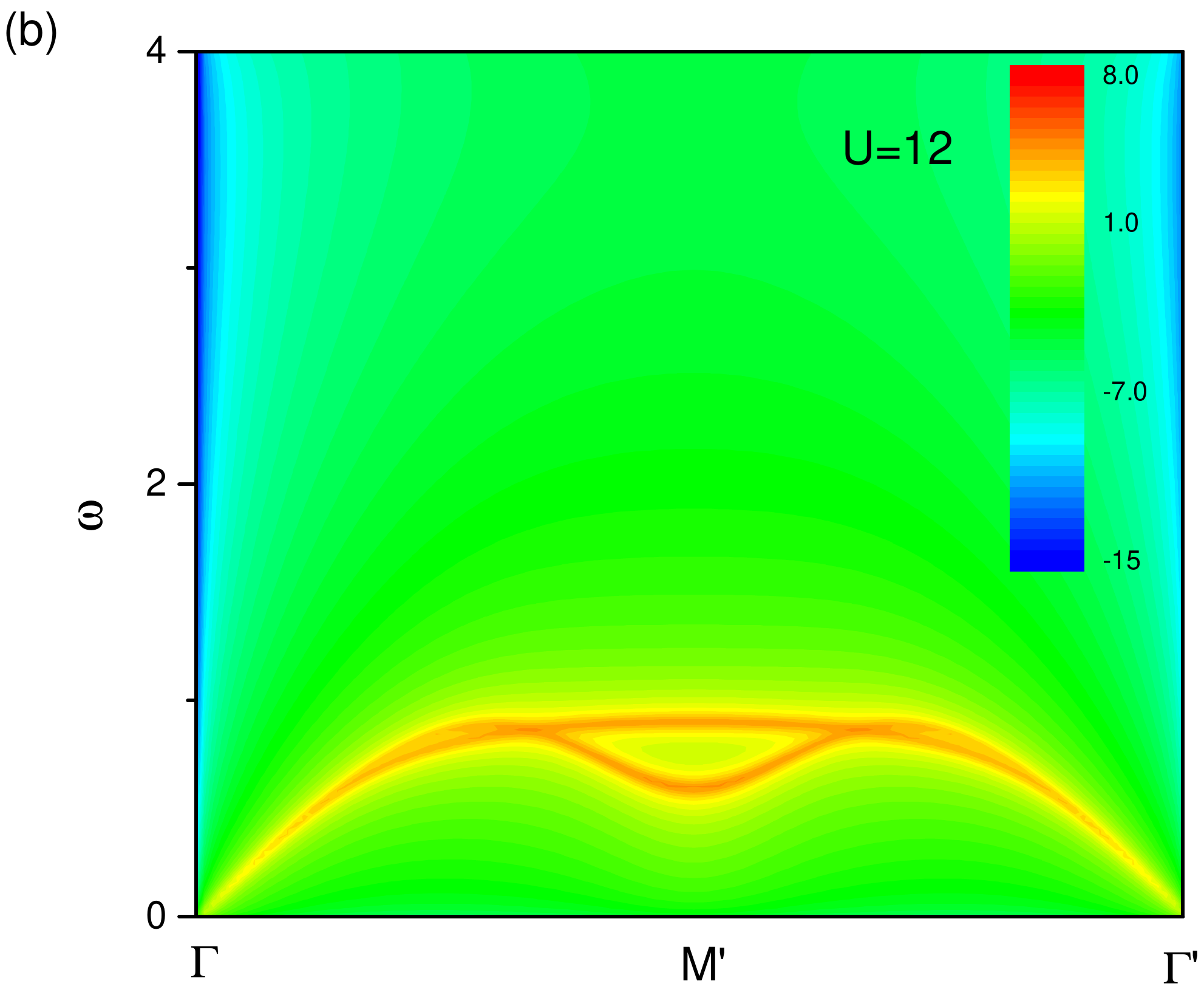}
\includegraphics[width=4.2cm,height=3cm]{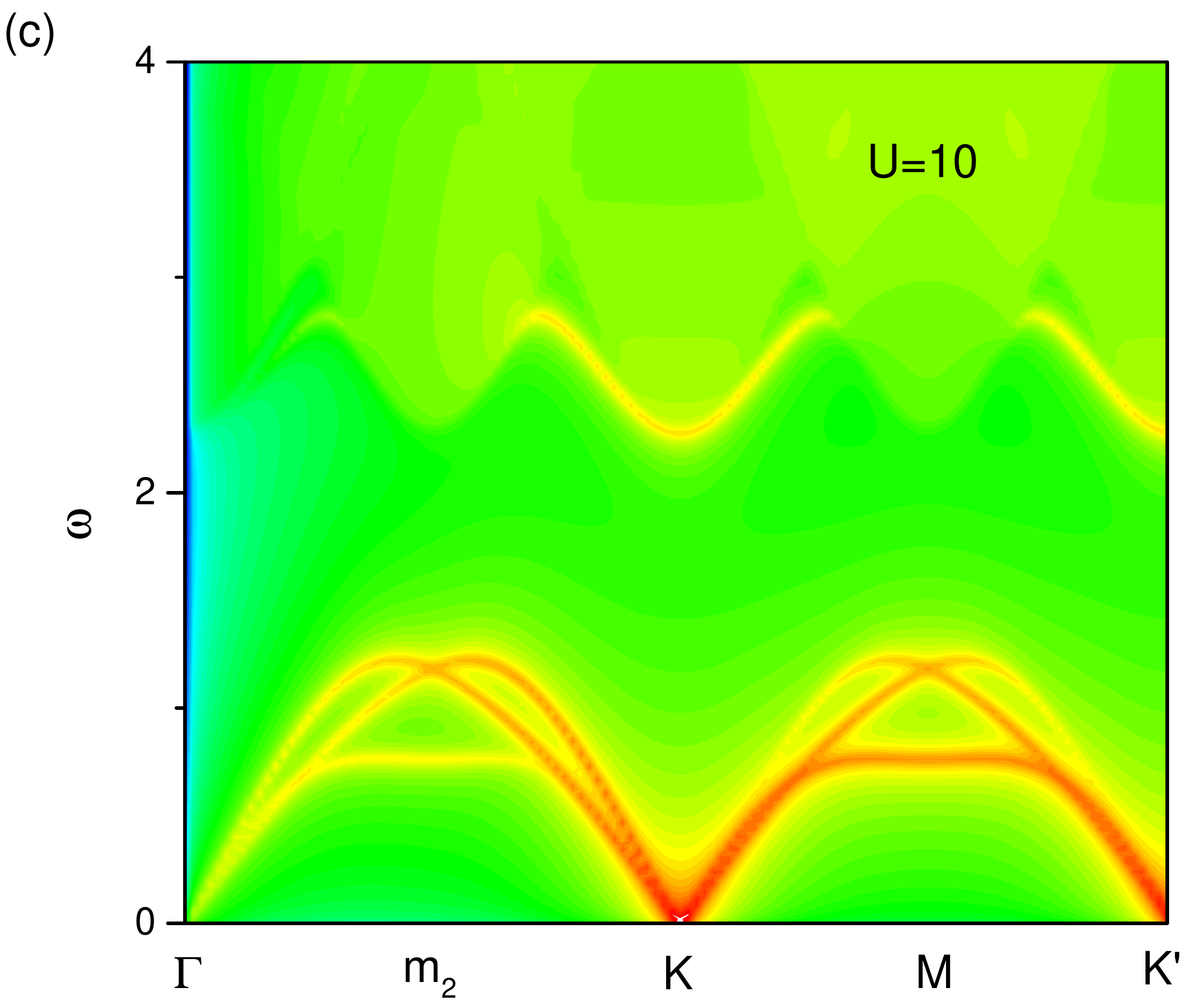}
\includegraphics[width=4.2cm,height=3cm]{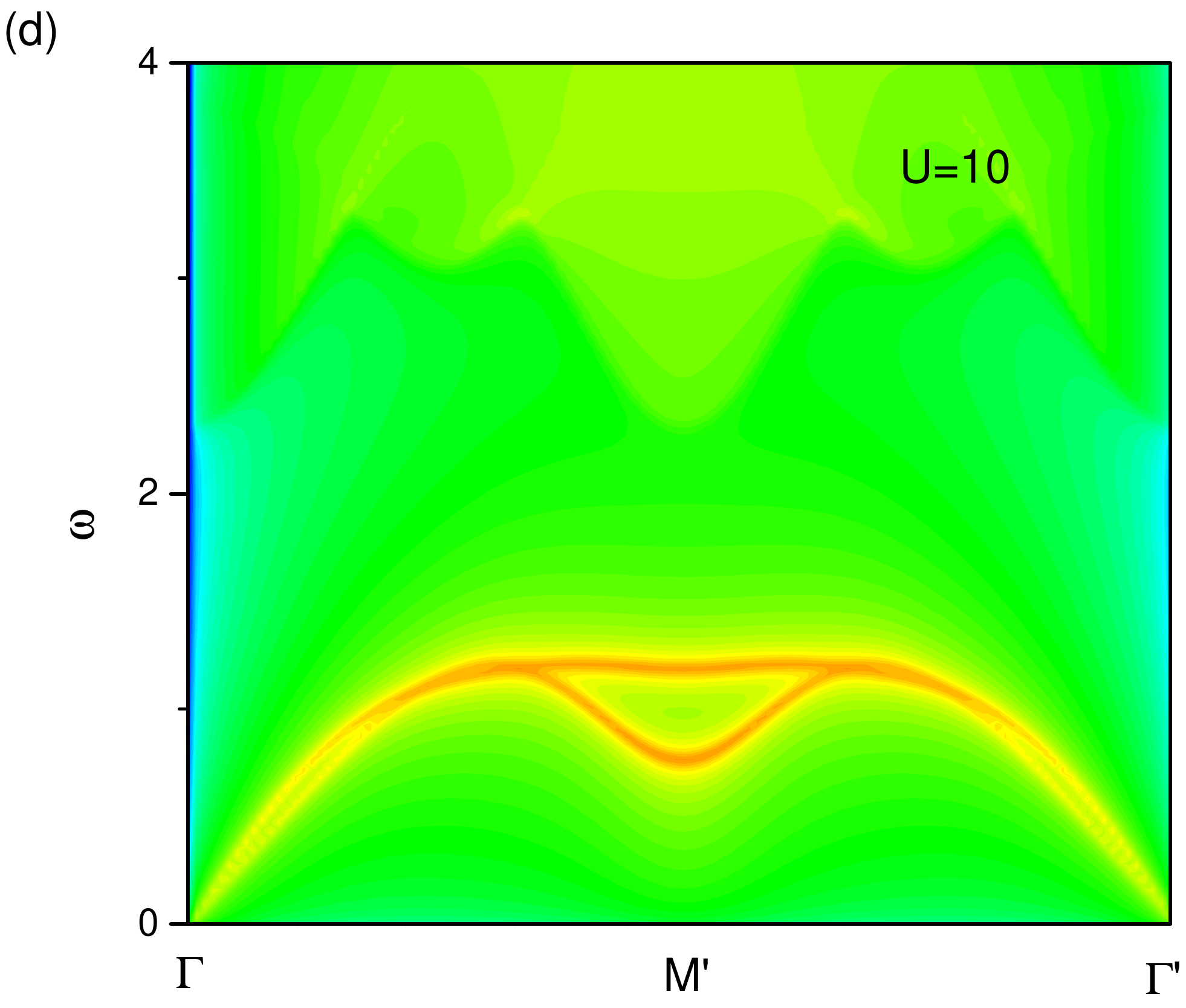}
\includegraphics[width=4.2cm,height=3cm]{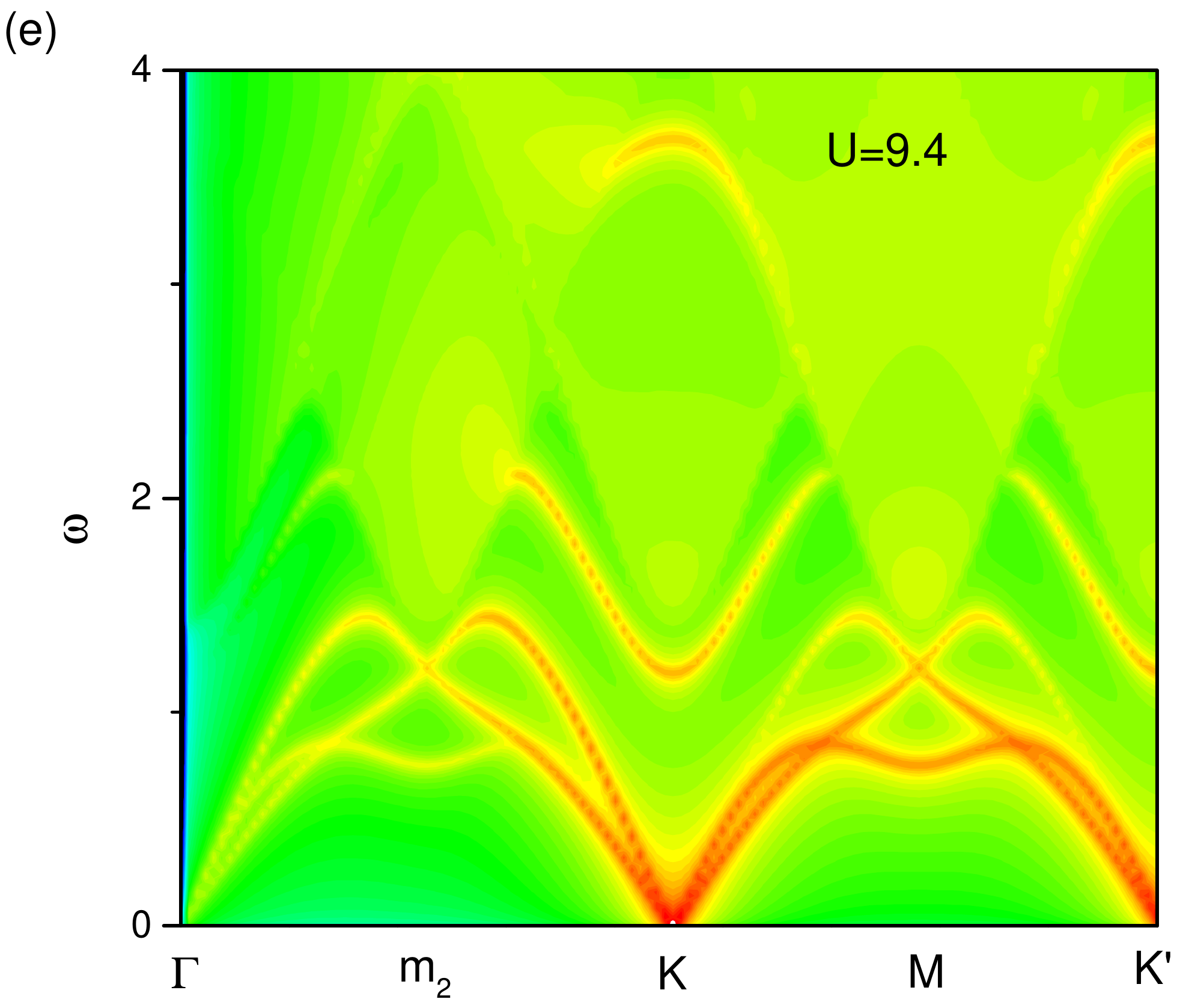}
\includegraphics[width=4.2cm,height=3cm]{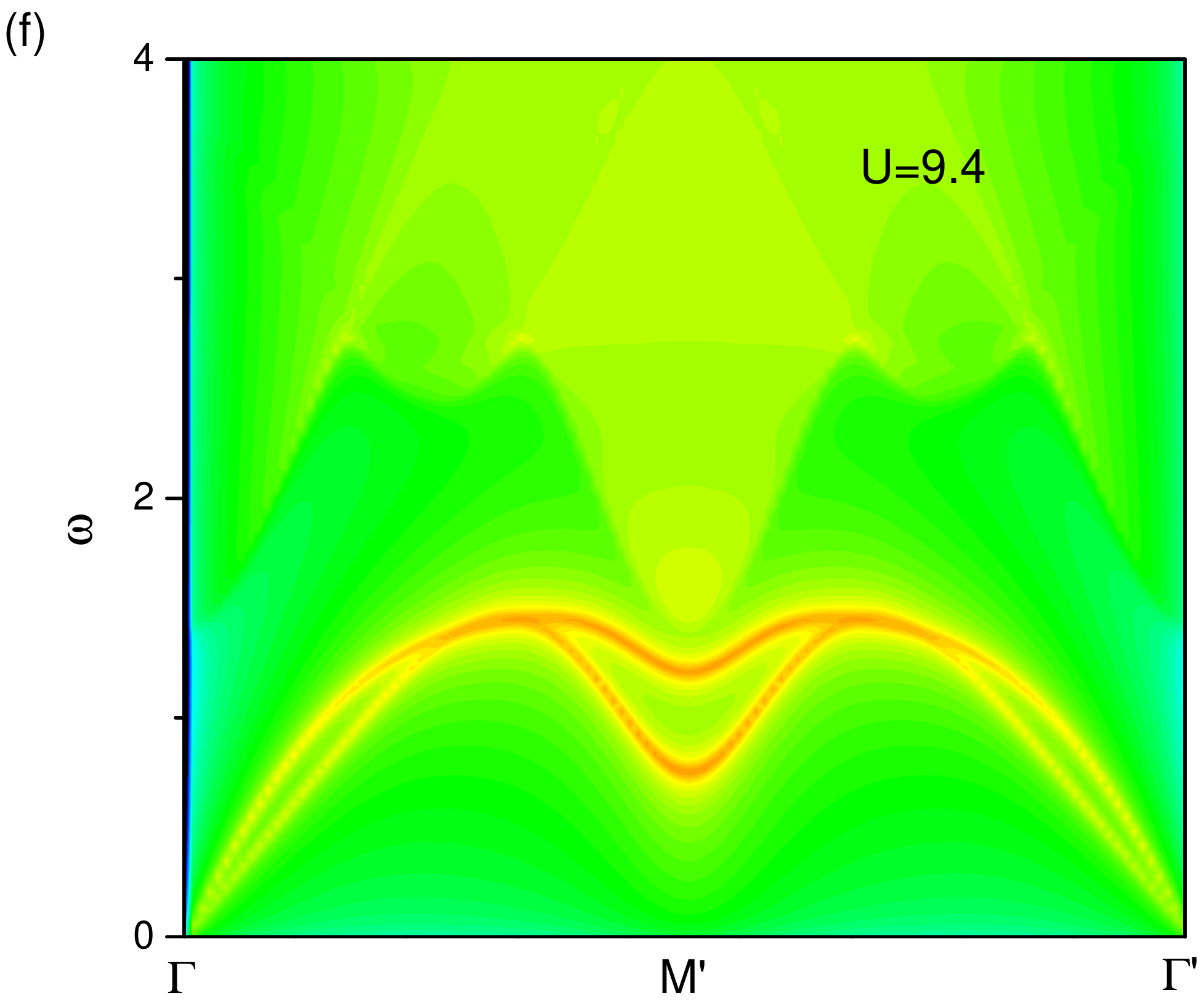}
\includegraphics[width=4.2cm,height=3cm]{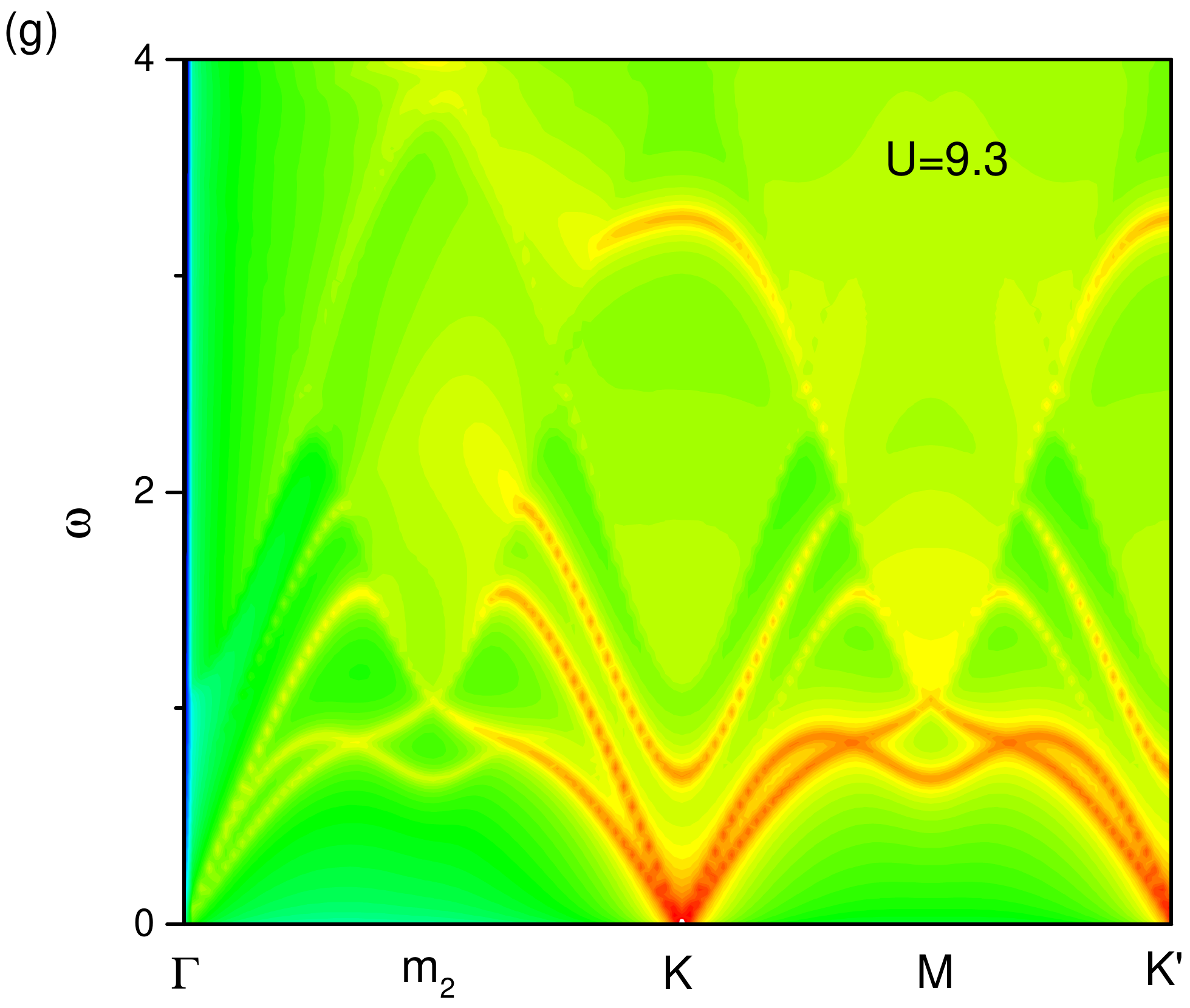}
\includegraphics[width=4.2cm,height=3cm]{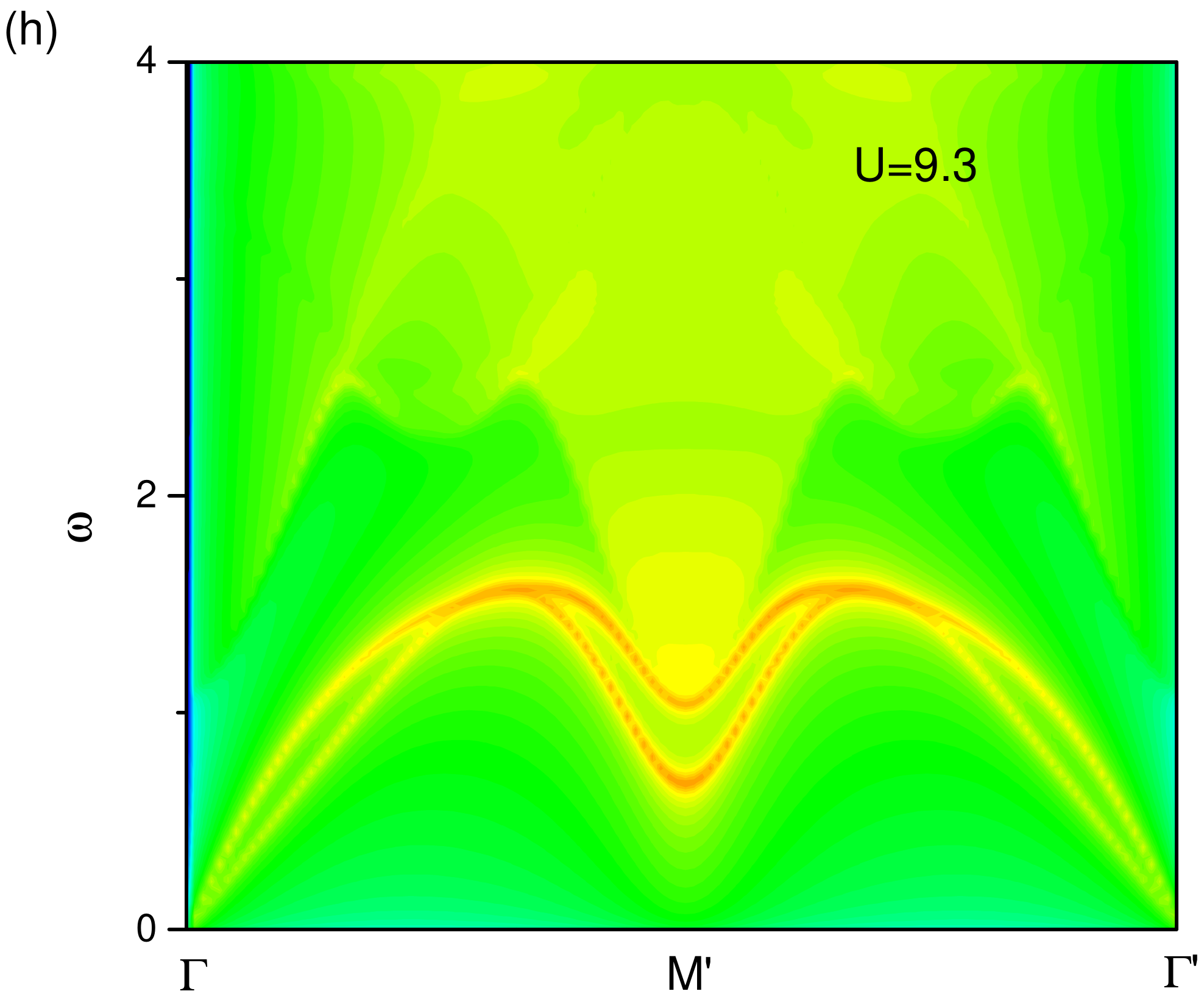}
\includegraphics[width=4.2cm,height=3cm]{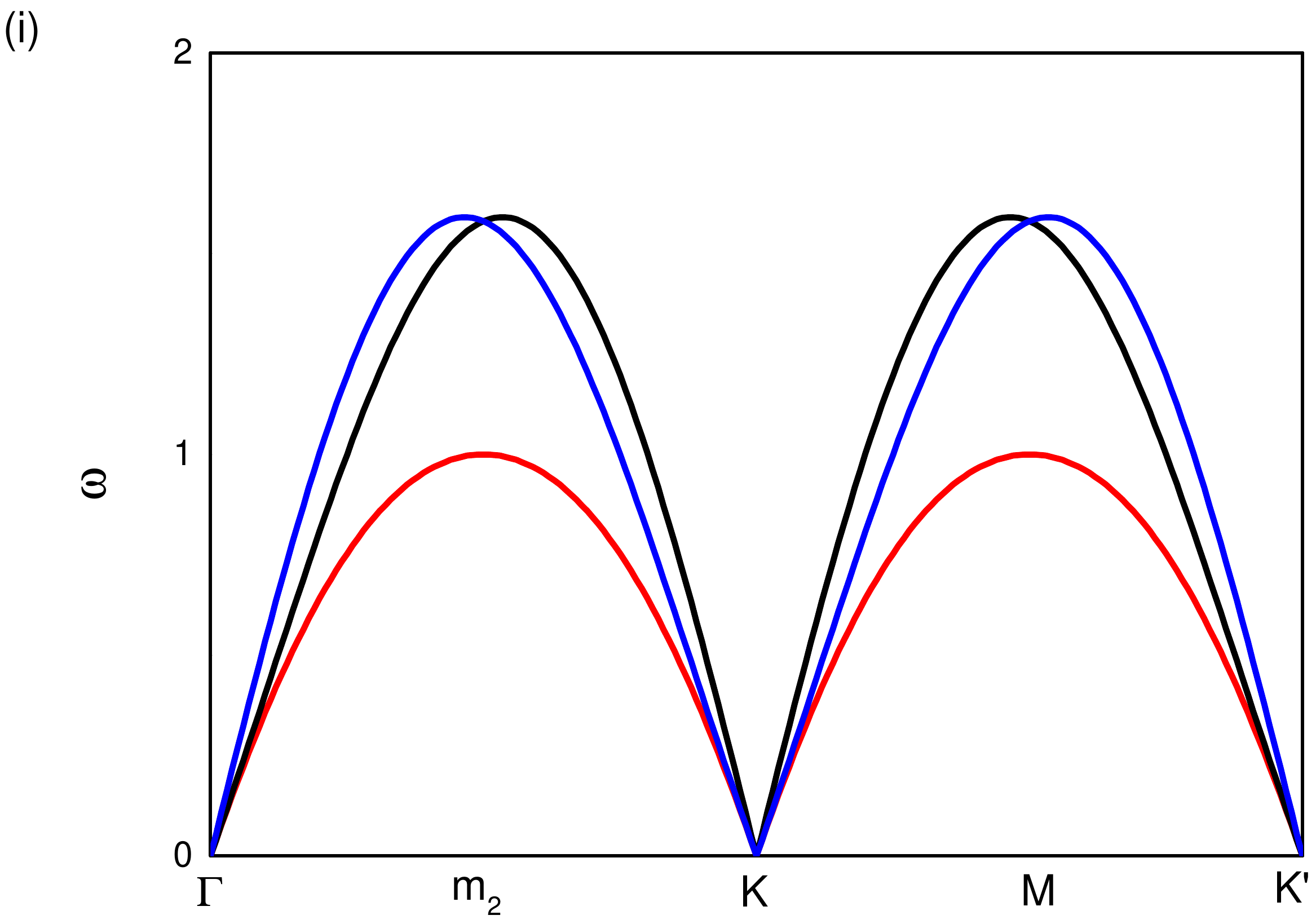}
\includegraphics[width=4.2cm,height=3cm]{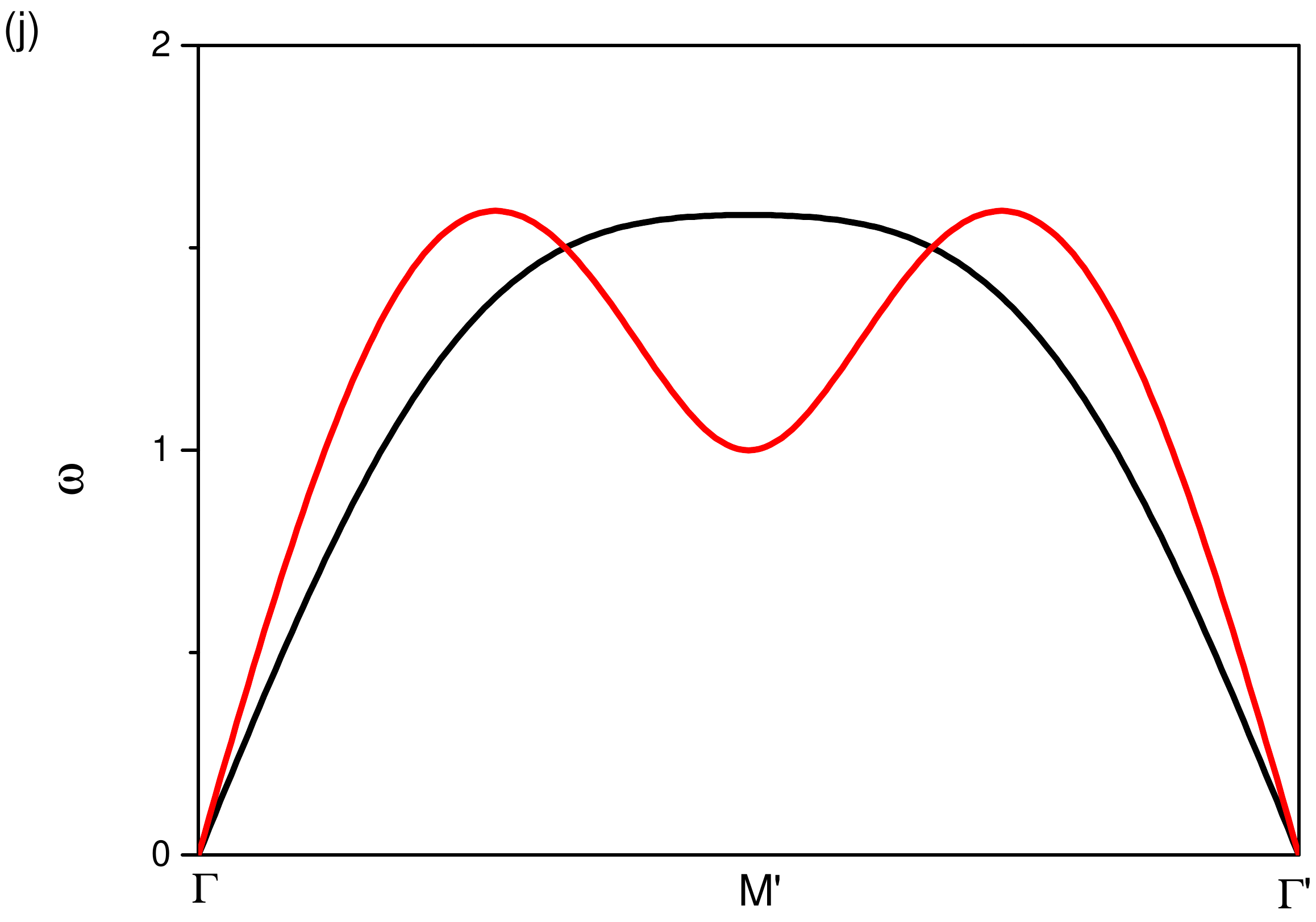}
\caption{The evolution of the RPA-corrected spin fluctuation spectrum with $U$ along $\Gamma-\mathbf{K}-\mathbf{K'}$(left column) and $\Gamma-\mathbf{M}'-\Gamma'$(right column). (a) and (b)$U$=15. (c) and (d)$U$=10. (e) and (f)$U$=9.4. (g) and (h)For $U$=9.3, which is very close to $U_{c}$. (i) and (j)The magnon dispersion predicted by the LSWT. Here the spectral intensity is plotted in logarithmic scale for clarity. The color scale shown in (b) is used in all plots of this figure and the following figures of this paper.}
\end{figure}

\begin{figure}
\includegraphics[width=4.2cm,height=3cm]{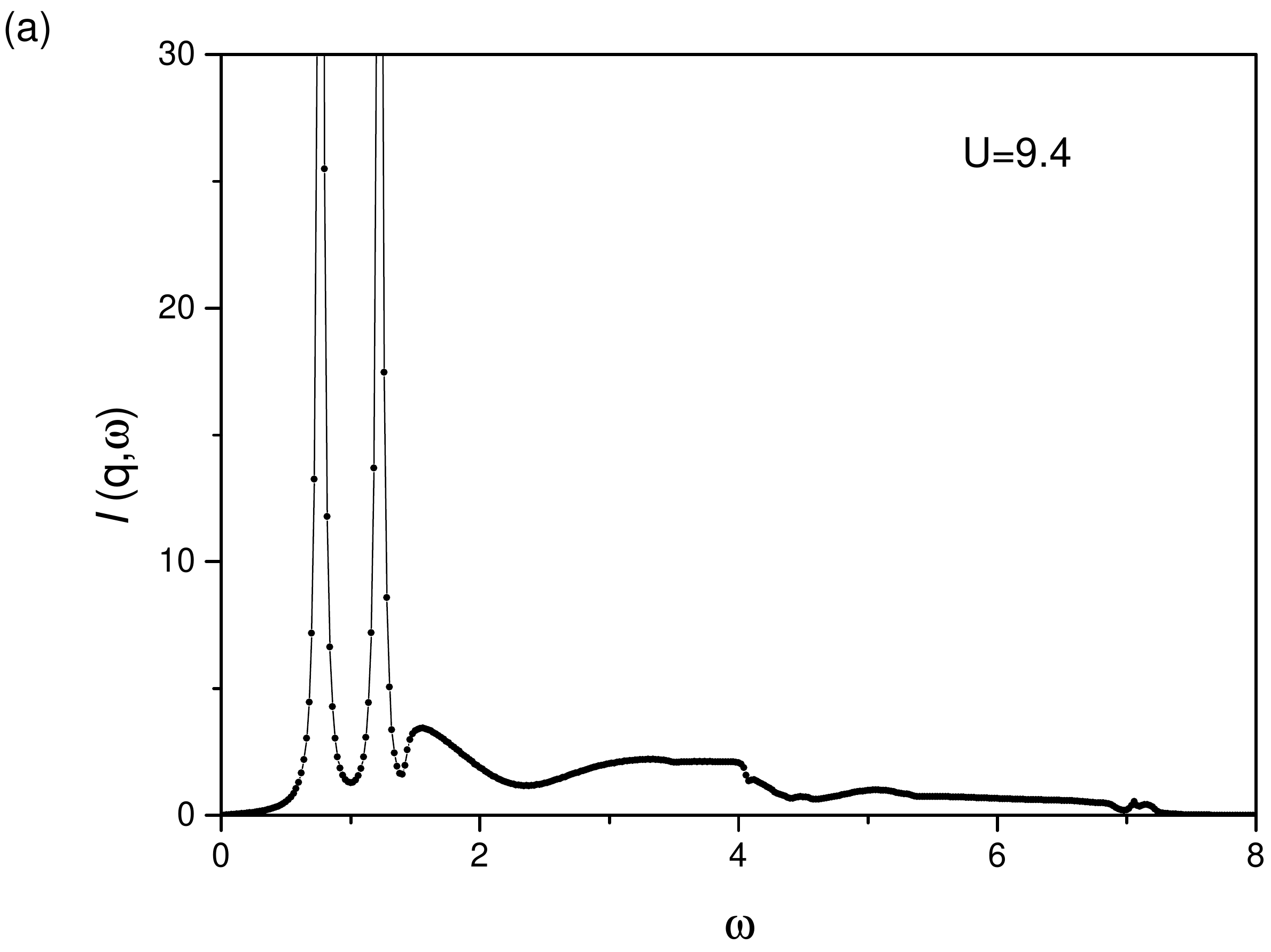}
\includegraphics[width=4.2cm,height=3cm]{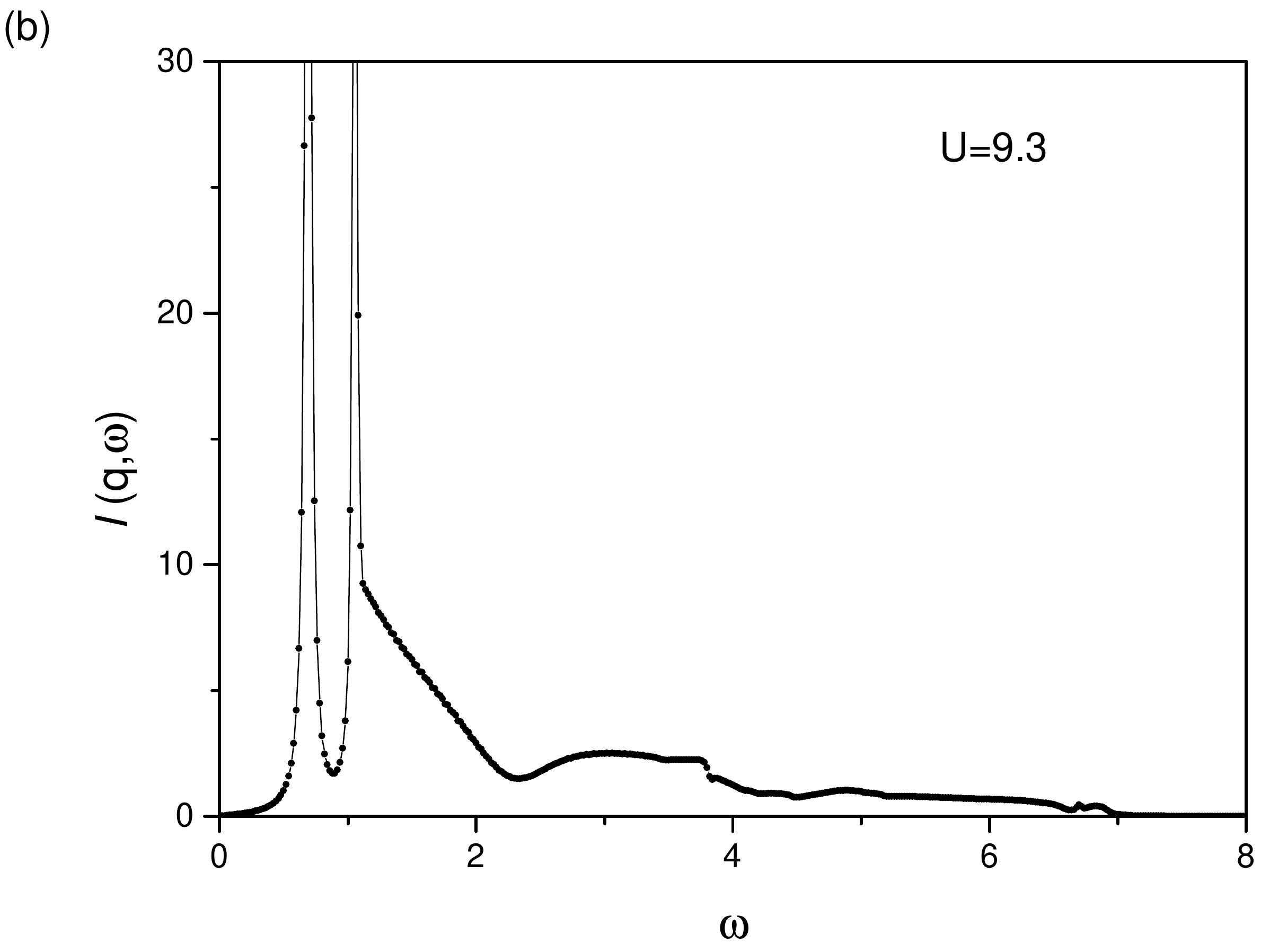}
\caption{The spin fluctuation spectrum at the M point for $U$=9.4(left) and $U$=9.3(right). The transfer of spectral weight from the magnon mode to the continuum is evident.}
\end{figure}

When we decrease $U$ further toward $U_{c}$, the downward renormalization of the magnon dispersion become even more significant. The intensity of the fourth collective mode around the $\mathbf{K}$ point is also significantly enhanced. As is shown in Fig3e for the case of $U$=9.4, the upper two magnon branches almost touch the bottom of the spinon continuum at the M point and start to transfer spectral weight into the continuum. Indeed, a broad peak start to develop close to the bottom of the spinon continuum(see Fig.4a). The broad peak is separated from the magnon mode below the continuum by a dip in the spectral function. When $U$ is very close to $U_{c}$, for example, in the case of $U$=9.3, the renormalization becomes so strong that magnon dispersion far away from the M point is also strongly modified. The dip between magnon peak and the broad peak in the continuum is now also filled up(see Fig.4b).

The three branches of magnon mode in the MBZ can actually be understood as those folded from a single magnon branch in the lattice Brillouin zone by three different magnetic reciprocal vectors. For example, the three magnon branches along $\mathbf{K}-\mathbf{K'}$ can be understood as those folded from $\mathbf{K}_{3}-\mathbf{K}_{2}$, $\Gamma-\mathbf{K}$ and $\mathbf{K}_{4}-\Gamma$ respectively.  In the 120 degree ordered phase of the spin-$\frac{1}{2}$ TLHAF, the system Hamiltonian is invariant under the combined operation of lattice translation and spin rotation by 120 degree in the ordering plane. The Hamiltonian thus takes a translational invariant form if a local coordinate system with its x-axis aligned to the ordered moment and its z-axis perpendicular to the ordering plane is adopted. Such a generalized translational symmetry prohibits the magnon mode folded from the $\mathbf{m}_{1}$ and the $\mathbf{m}_{2}$ point to hybridize at the M point of the MBZ. We note that as a result of the $\pi$-flux structure in the mean field Hamiltonian, the spinon system is invariant under such a generalized translation operation only up to a gauge transformation. However, since the spin operator is a gauge invariant object, the spin dynamics calculated from the spinon theory is still invariant under such a generalized translational symmetry.

The evolution of the RPA-corrected spin fluctuation spectrum with $U$, especially, the strong deviation of the magnon dispersion around the M point for $U$ close to $U_{c}$, can be understood from the level repulsion effect between the spinon continuum and the collective spin excitation below it. As we discussed above, the most important feature of the spinon continuum in the $\pi$-flux phase is the Dirac cone structure around the M point, which results from the particle-hole excitation of the spinon system between its two Dirac nodes. Such a strongly momentum dependent level repulsion effect enhances with the decrease of $U$ and is responsible for the formation of the roton-like minimum in the magnon dispersion around the M point. In this picture, we always expect the crossing of the upper two magnon branches at the M point, which is protected by the generalized translational symmetry mentioned above. In fact, this crossing should be expected in any correct theory of the spin dynamics of the spin-$\frac{1}{2}$ TLHAF. As a result, these two modes should either both show up around the M point, or dissolve into the continuum together. In other words, we should never expect just two branches of magnon mode to show up around the M point. This general conclusion from the symmetry analysis seems to be in conflict with the INS result on Ba$_{3}$CoSb$_{2}$O$_{9}$.
   
The fourth collective mode around the $\mathbf{K}$ point can be attributed to longitudinal spin fluctuation in the 120 degree ordered phase, as can be verified by its polarization character below. In particular, it reduces to the fluctuation in the amplitude of the magnetic order parameter at the $\mathbf{K}$ point. The spectral weight of this mode is drawn from the spinon continuum around the $\mathbf{K}$ point, which, according to discussion in the last subsection, is caused by the particle-hole transition of the spinon system within the same Dirac node. This understanding is consistent with the fact that the fourth mode moves up with the increase of U and finally dissolves into the Dirac continuum, since the amplitude fluctuation of the ordered moment becomes increasingly expensive in energy with the increase of the interaction strength.

Beside the four collective modes discussed above, one also observe some spectral feature inside the spinon continuum at intermediate energies. These spectral features are caused by Van Hove singularities in the joint density of the state(JDOS) of the spinon excitation. However, we note that the prediction on these high energy spectral features from the RPA theory are much less reliable than those on the collective modes, since the sharp features of JDOS inside the spinon continuum are not robust against the spinon self-energy correction, which are neglected in the RPA treatment here.

To summarize, the RPA theory predicts that there are three branches of magnon mode and one longitudinal mode in the MBZ. There are also strong spin fluctuation continuum around the M point at higher energy. The level repulsion effect between the spinon continuum and the magnon mode can generate strongly momentum dependent renormalization of the magnon dispersion and is responsible for the roton-like minimum around the M point. However, the momentum range in which the magnon dispersion is strongly modified seems to be too broad to account for the INS result. 

To improve the agreement with the INS result, we now study the RPA-corrected spin fluctuation spectrum with both the Hubbard and the Heisenberg interaction. As we discussed before, the Hubbard interaction can be understood as a first approximation to spinon correlation. The spinon can also interact in the spin density channel through the Heisenberg term. However, since the Heisenberg term has already been decoupled in the RVB channel in the spinon description, we must understand the Heisenberg interaction between the spinon in the spin density channel as a residual interaction and treat its strength $J_{1}$ as a phenomenological parameter. 
   
In the 120 degree ordered phase, the Heisenberg term will contribute a mean field of strength $3J_{1}|\mathbf{m}|$ in the spin density channel. The RPA kernel is also modified to be $\mathbf{V}(\mathbf{q})=\frac{4U}{3}\mathbf{I}-\mathbf{J}(\mathbf{q})$, with $\mathbf{J}(\mathbf{q})$ the Fourier transform of the Heisenberg interaction. $\mathbf{J}(\mathbf{q})$ is a $18\times18$ matrix with the following matrix elements
\begin{equation}
J^{i,j}_{\mu,\nu}(\mathbf{q})=J_{1}\delta_{i,j}M_{\mu,\nu},
\end{equation}  
in which $M_{\mu,\nu}$ denotes the matrix element of the following $6\times6$ matrix
\begin{equation}
\mathbf{M}=\left(\begin{array}{cccccc}0 & \gamma_{1} & \gamma_{2} & 0 &  \gamma_{3} & \gamma_{4}\\
                                                            \gamma^{*}_{1} & 0 & \gamma_{1} &1 & 0 & \gamma_{3}\\
                                                            \gamma^{*}_{2} &  \gamma^{*}_{1} & 0&\gamma^{*}_{1} & 1 & 0 \\
                                                            0 & 1 & \gamma_{1} &0 & \gamma_{1} & \gamma_{2}\\
                                                             \gamma^{*}_{3}  & 0 & 1& \gamma^{*}_{1} & 0 & \gamma_{1} \\
                                                             \gamma^{*}_{4}  &  \gamma^{*}_{3} & 0 & \gamma^{*}_{2} & \gamma^{*}_{1} & 0\end{array}\right).
\end{equation}
Here $\gamma_{1}=1+e^{iq_{2}}$, $\gamma_{2}=e^{iq_{2}}$, $\gamma_{3}=e^{iq_{1}}$, $\gamma_{4}=e^{iq_{1}}(1+e^{iq_{2}})$.

\begin{figure}
\includegraphics[width=4.2cm,height=3cm]{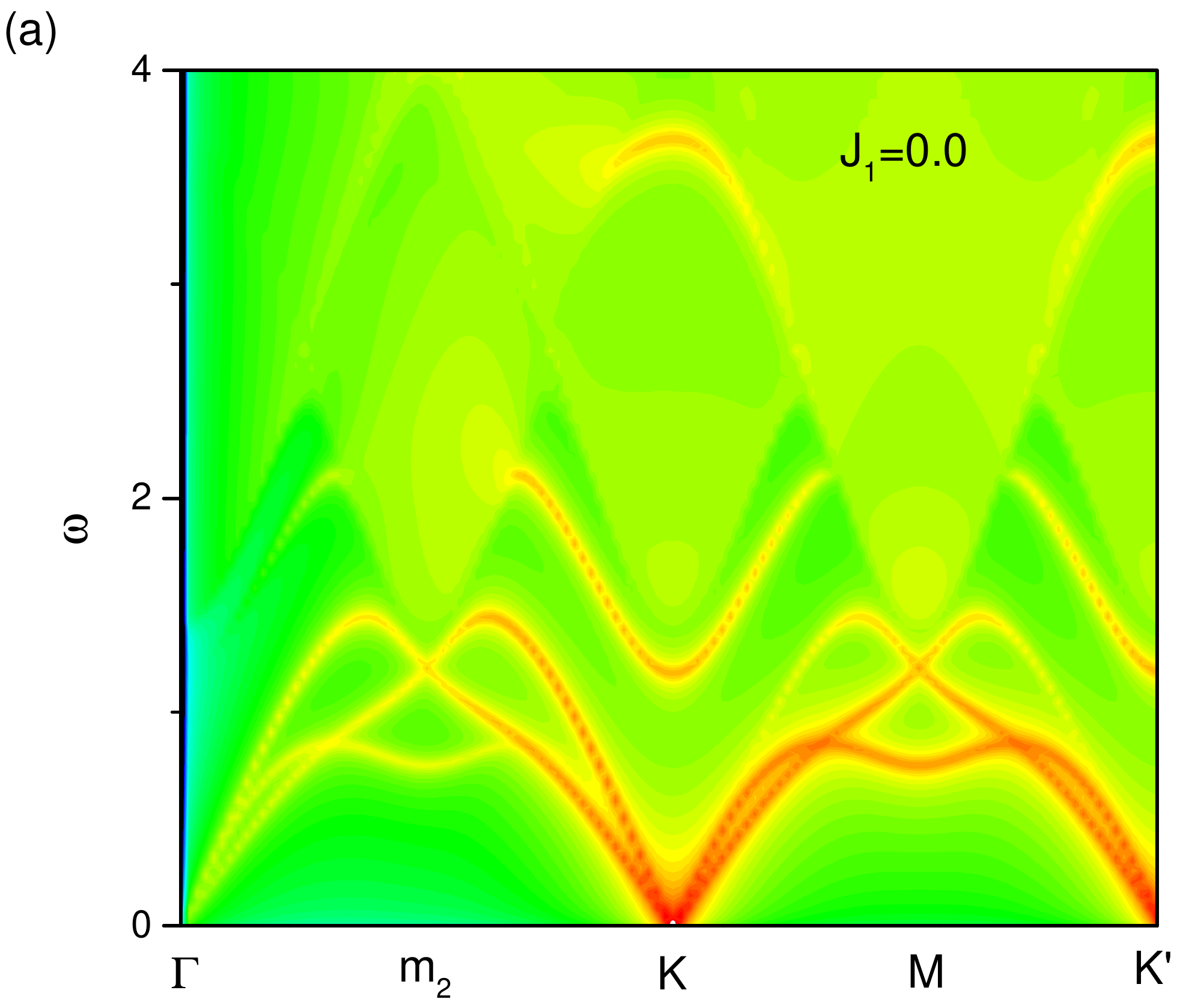}
\includegraphics[width=4.2cm,height=3cm]{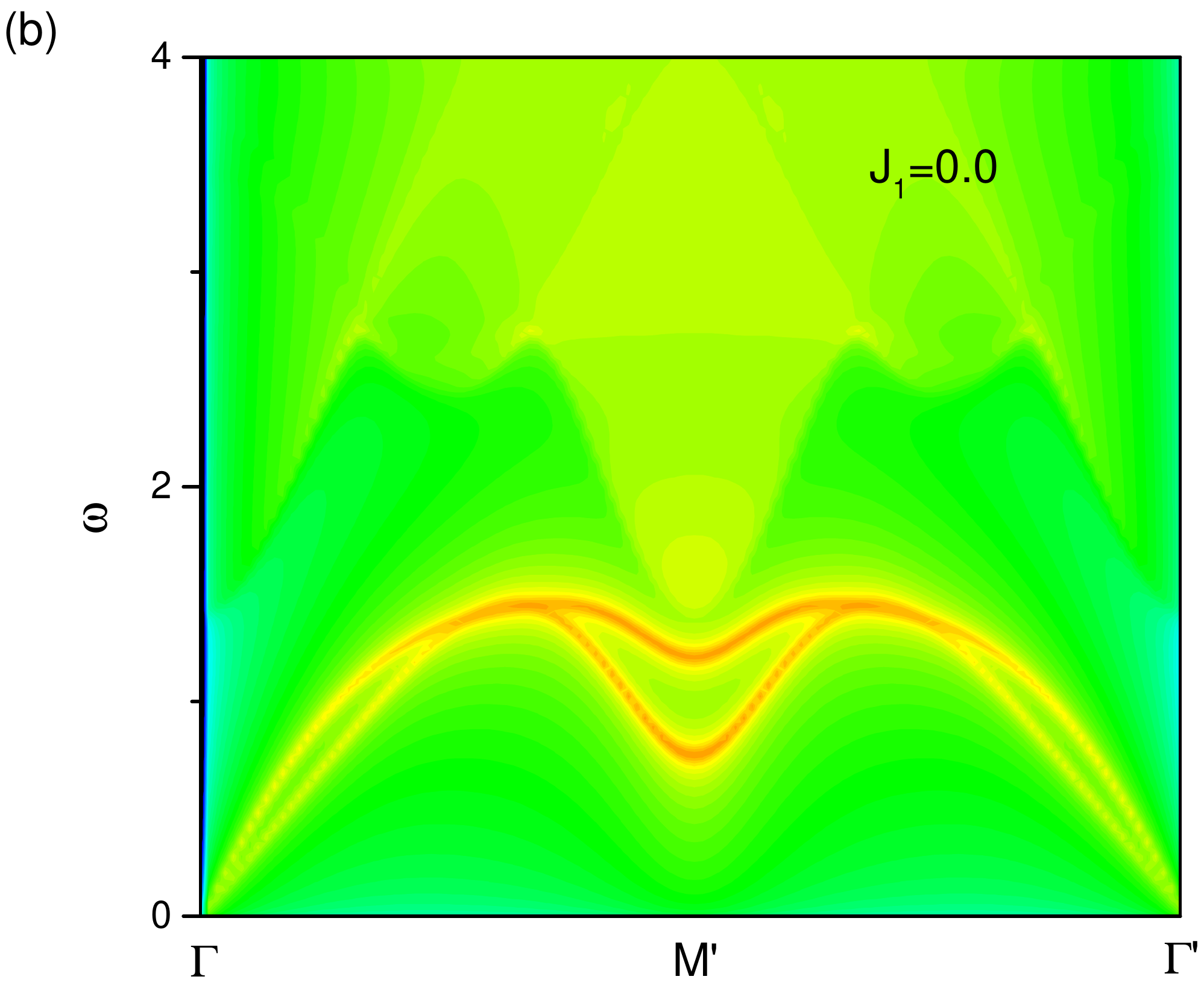}
\includegraphics[width=4.2cm,height=3cm]{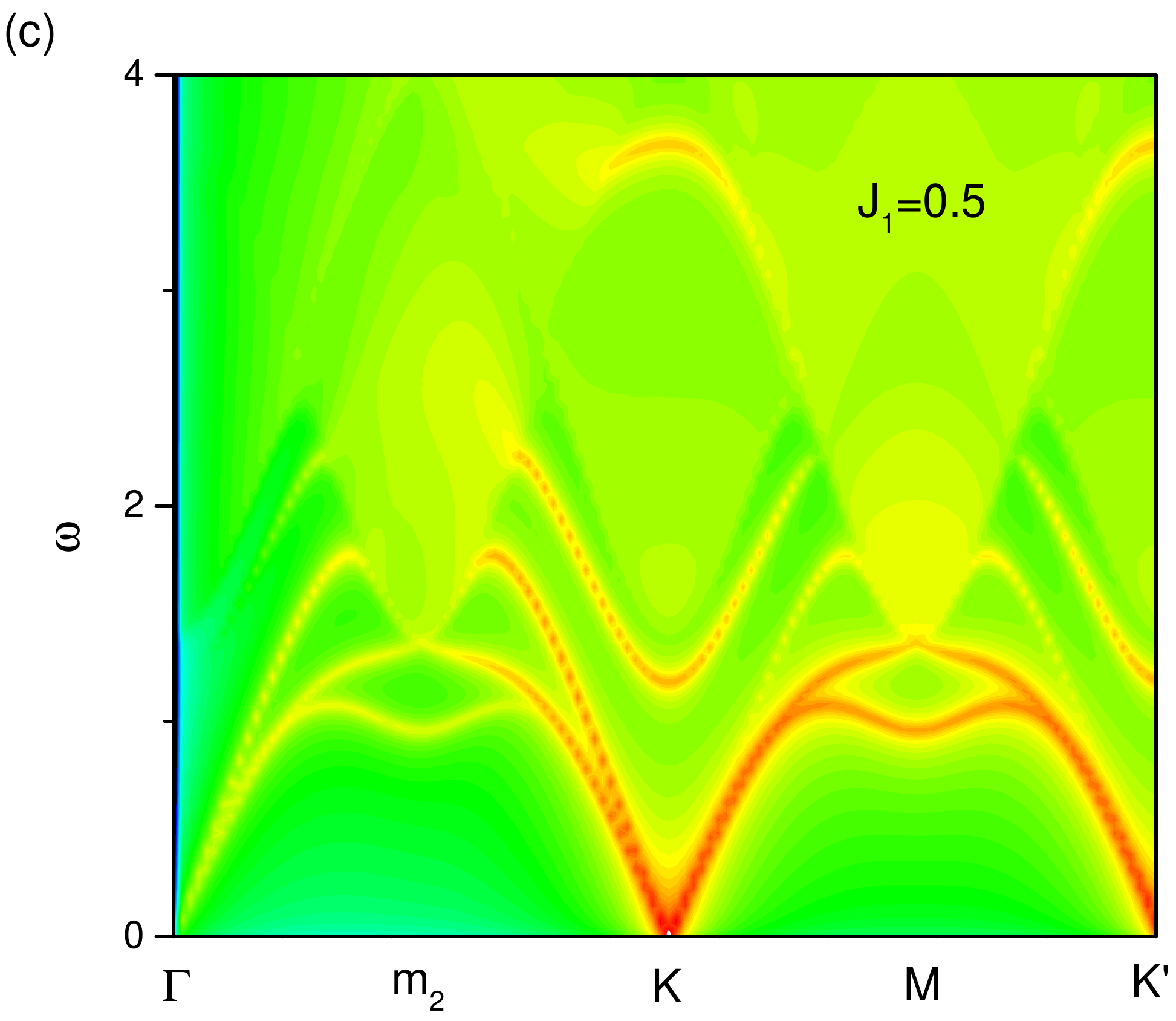}
\includegraphics[width=4.2cm,height=3cm]{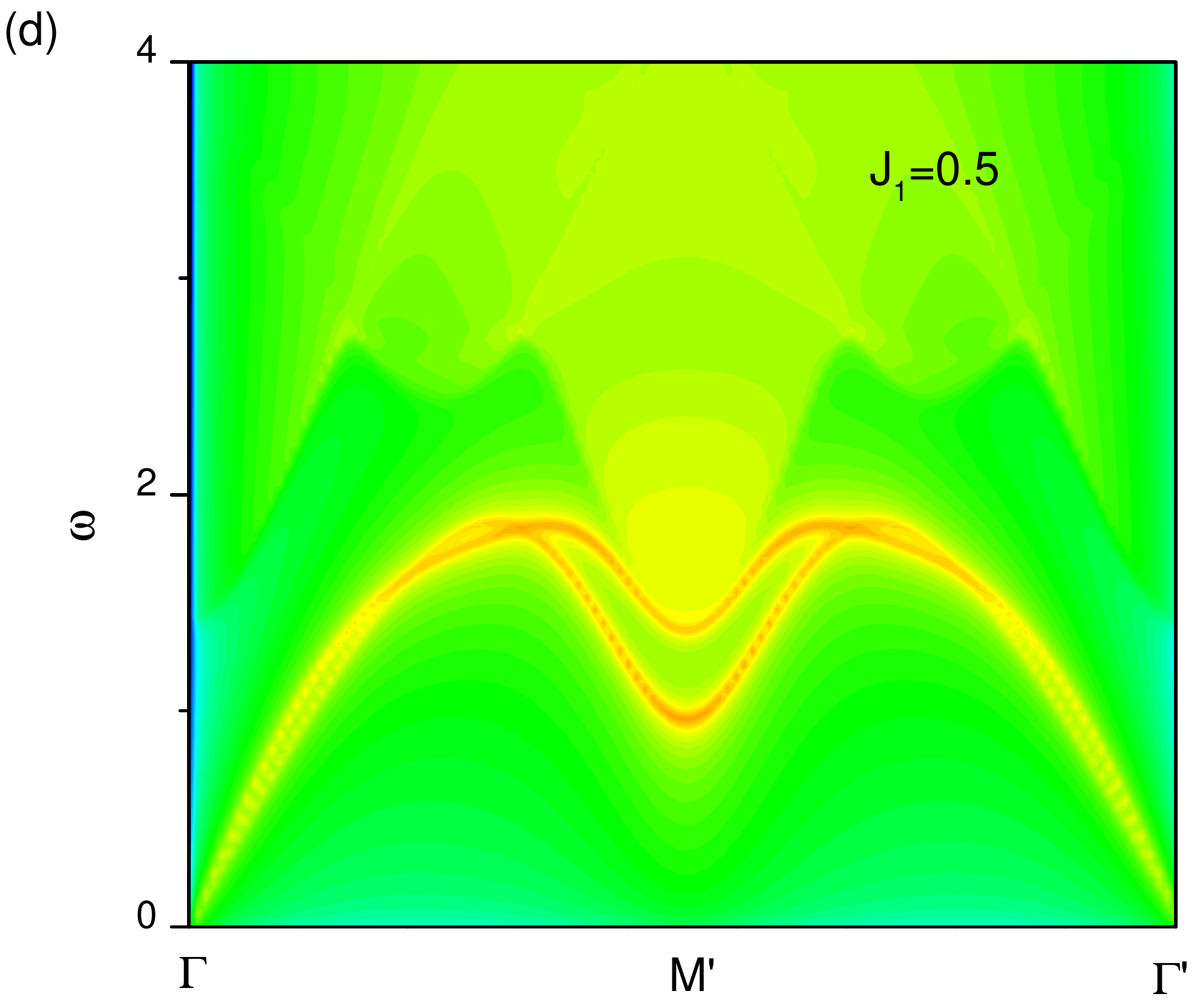}
\includegraphics[width=4.2cm,height=3cm]{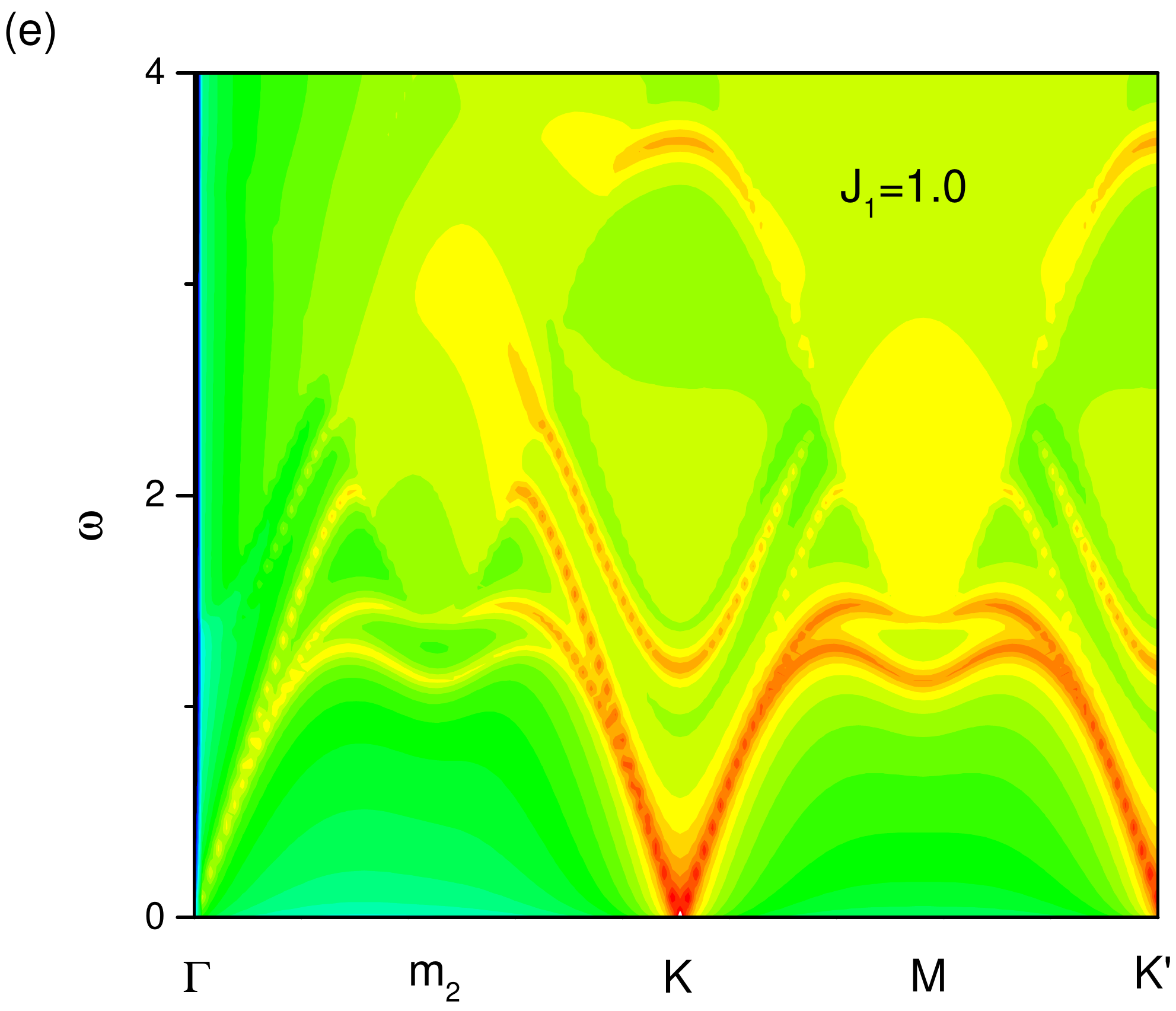}
\includegraphics[width=4.2cm,height=3cm]{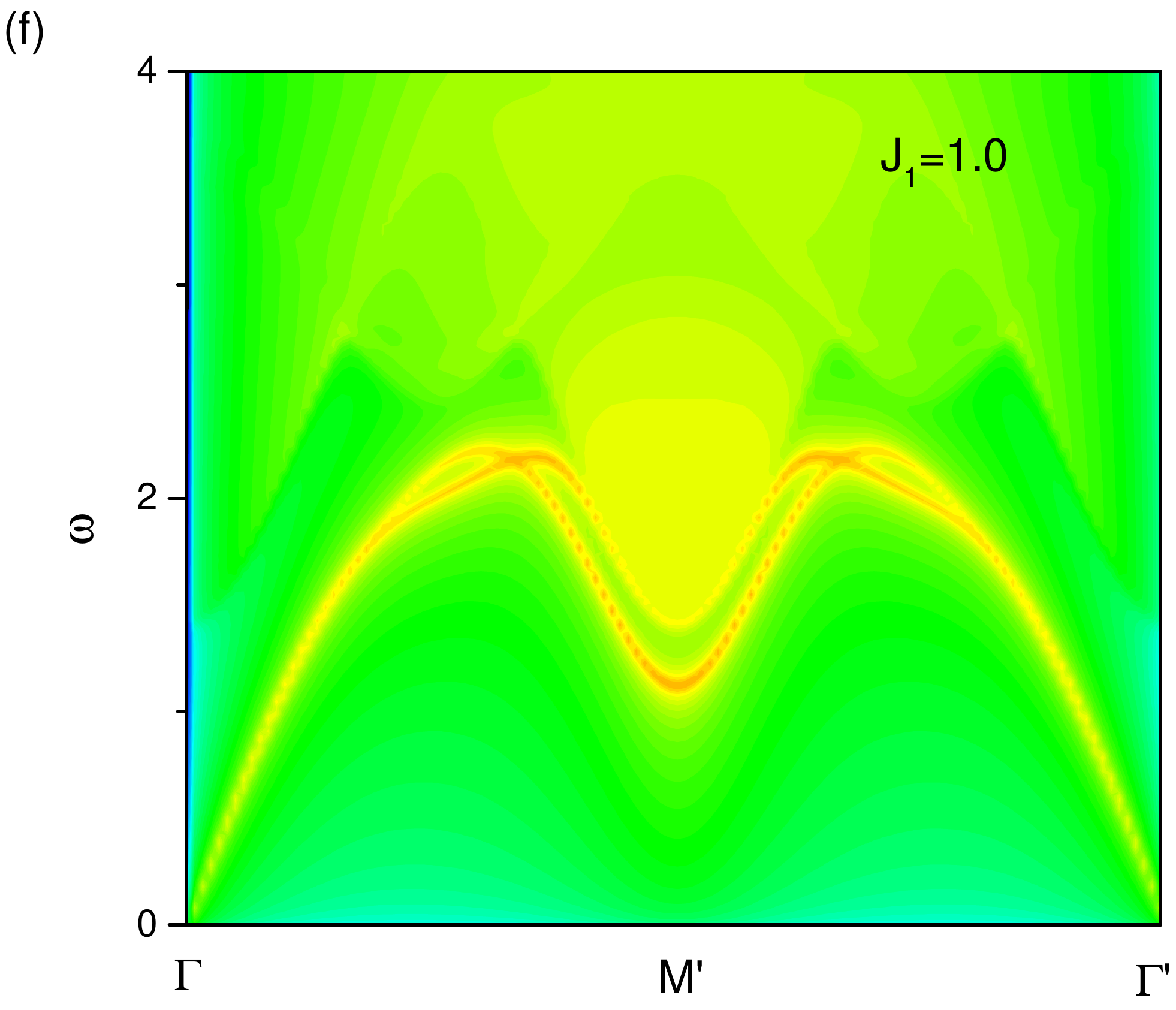}
\caption{The spin fluctuation spectrum along $\Gamma-\mathbf{K}-\mathbf{K}'$(left) and $\Gamma-\mathbf{M}'-\Gamma'$(right) calculated with both the Hubbard and the Heisenberg interaction in the RPA correction. Plotted here is the spectrum for $J_{1}$=0, 0.5 and 1. The spectral intensity is plotted in logarithmic scale for clarity. The same color scale as that shown in Fig.3b is used in all plots of this figure.}
\end{figure}

To make possible a meaningful comparison between the results calculated with different parameters, we keep the value of $(\frac{4U}{3}+3J_{1})$, and thus the strength of the mean field in the spin density channel fixed in the following calculation. More specifically, we choose $(\frac{4U}{3}+3J_{1})=\frac{4U^{*}}{3}$ with $U^{*}=9.4$ and tune the value of $J_{1}$. In Fig.5 we plot the RPA-corrected spin fluctuation spectrum with different values of $J_{1}$. The spectrum with both the Hubbard and the Heisenberg interaction are similar to those with only the Hubbard interaction.  However, the momentum range in which the magnon dispersion is strongly modified is now much more restricted to the M point. 

In Ba$_{3}$CoSb$_{2}$O$_{9}$, the Heisenberg interaction is slightly anisotropic and the exchange coupling out of the ordering plane is about 5 percent smaller than that within the ordering plane. However, we note that such an easy plane anisotropy does not break the general translational symmetry we discussed above. So we expect that the spin fluctuation spectrum should not be changed qualitatively by such an exchange anisotropy. To check on this point, we have calculated the RPA-corrected spin fluctuation spectrum with 
$J_{1}^{xx}=J_{1}^{yy}$=1 and $J_{1}^{zz}$=0.95. Here $J_{1}^{xx}$, $J_{1}^{yy}$ and $J_{1}^{zz}$ denote the exchange coupling between spin component in the $x$, $y$ and $z$ direction. The result is plotted in Fig.6. One find that the main effect of the easy plane anisotropy is to open a gap at $\Gamma$ and $\mathbf{K}$ point in the dispersion of the magnon with dominate $z$ polarization. Such a result can be understood trivially in the framework of the LSWT.

\begin{figure}
\includegraphics[width=4.2cm,height=3cm]{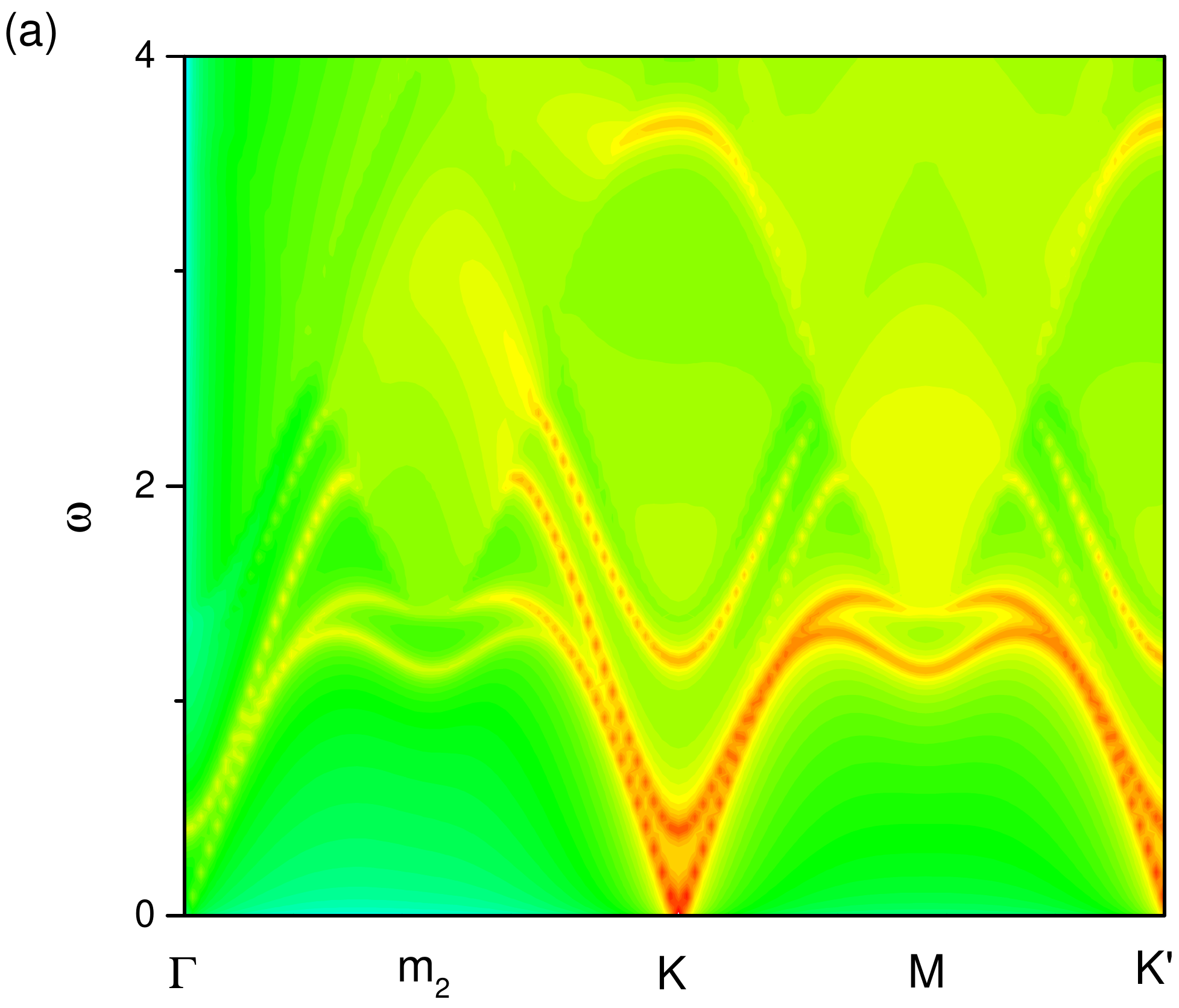}
\includegraphics[width=4.2cm,height=3cm]{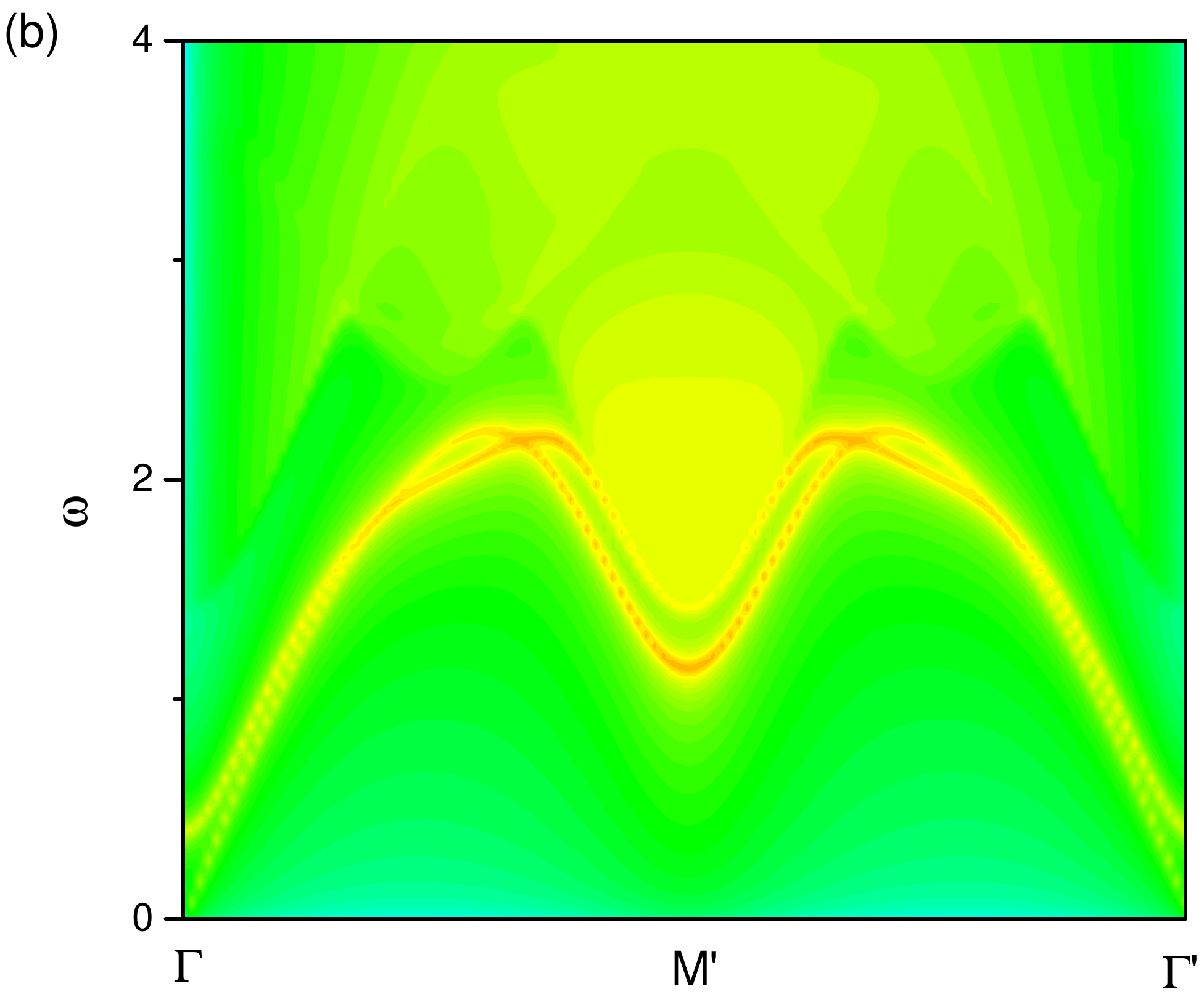}
\caption{The spin fluctuation spectrum along $\Gamma-\mathbf{K}-\mathbf{K}'$(left) and $\Gamma-\mathbf{M}'-\Gamma'$(right) calculated with both the Hubbard and an anisotropic Heisenberg interaction. Here $J_{1}^{xx}=J_{1}^{yy}$=1, $J_{1}^{zz}$=0.95. $U^{*}$=9.4. The spectral intensity is plotted in logarithmic scale for clarity. The same color scale as that shown in Fig.3b is used in all plots of this figure.}
\end{figure}

\subsection{Comparison with the INS results on Ba$_{3}$CoSb$_{2}$O$_{9}$}
In the INS measurement on Ba$_{3}$CoSb$_{2}$O$_{9}$, two coherent magnon modes are observed around the M point, whose dispersion are strongly renormalized to exhibit roton-like minimum around the M point. These two modes are accompanied by two strong and broad spectral peaks in the continuum at higher energy. Far away from the M point, four dispersive modes are observed, among which the two at lower energy are connected to the two coherent magnon modes at the M point. The other two modes at higher energy seem to be connected to the two broad spectral peaks in the continuum. However, according to the symmetry analysis presented in the last subsection, it is impossible to have just two branches of coherent magnon modes around the M point, since the crossing of the upper two magnon branch at the M point is protected by the generalized translational symmetry of the 120 degree ordered phase.

\begin{figure}
\includegraphics[width=4.2cm,height=3cm]{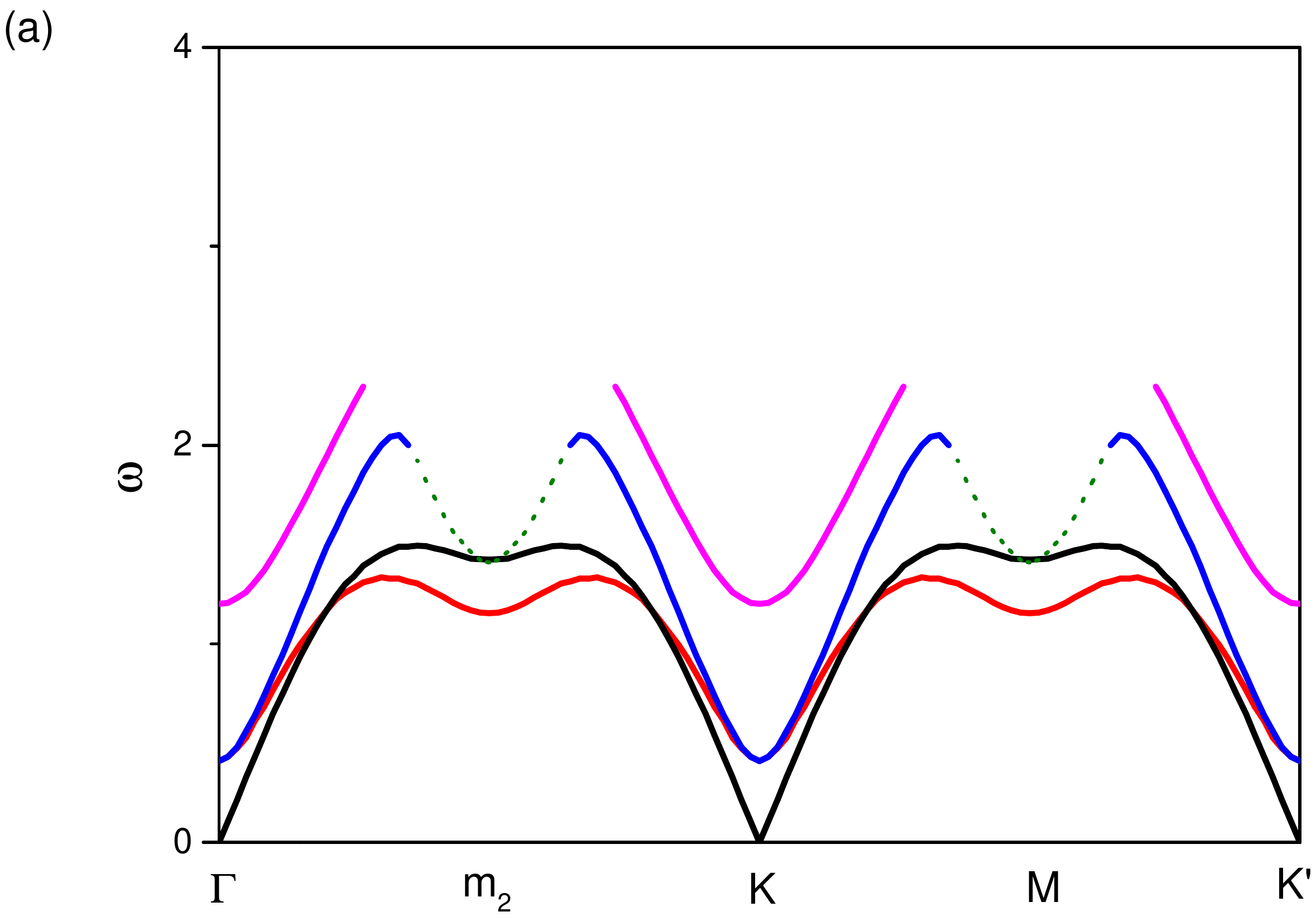}
\includegraphics[width=4.2cm,height=3cm]{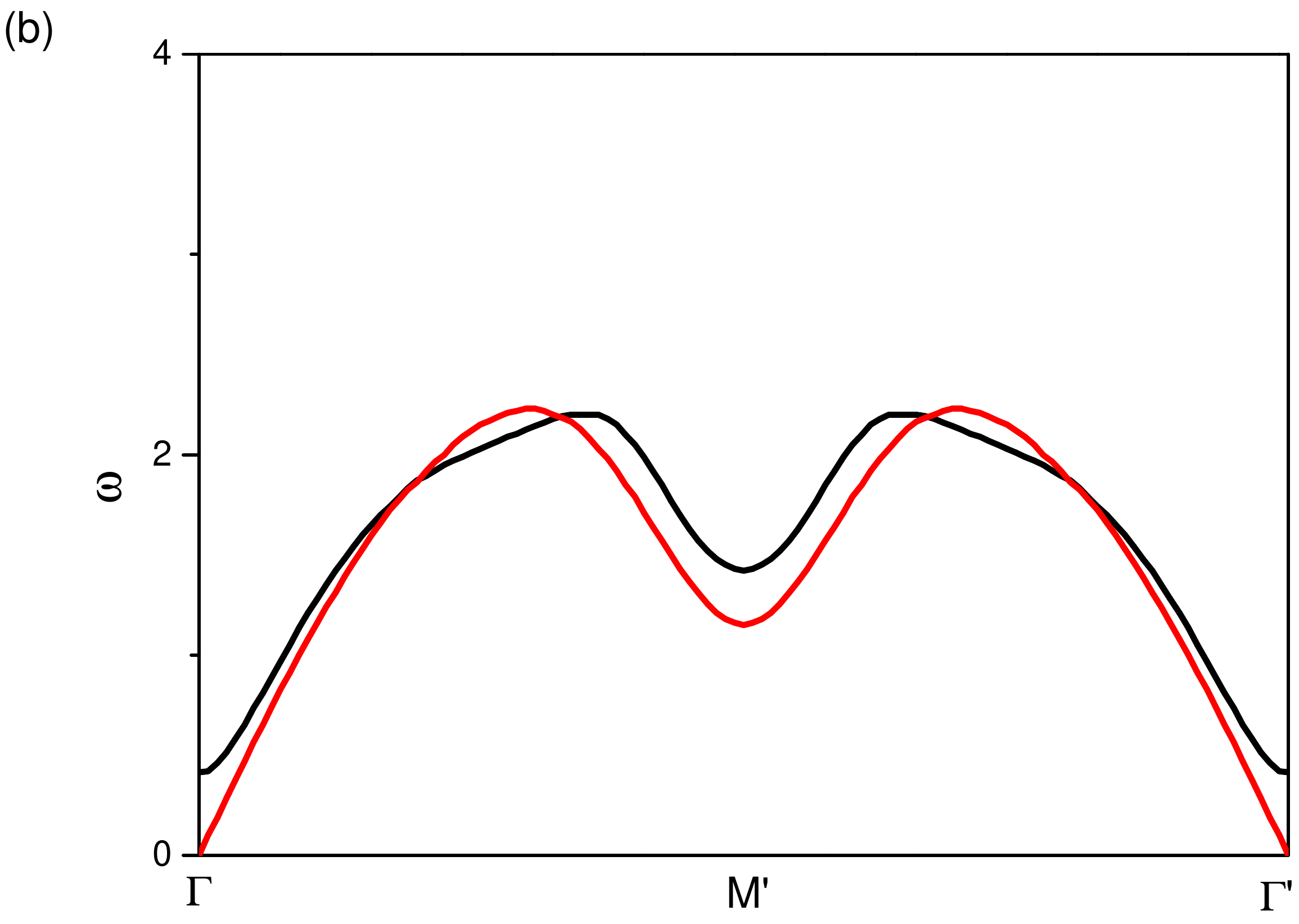}
\includegraphics[width=4.2cm,height=3cm]{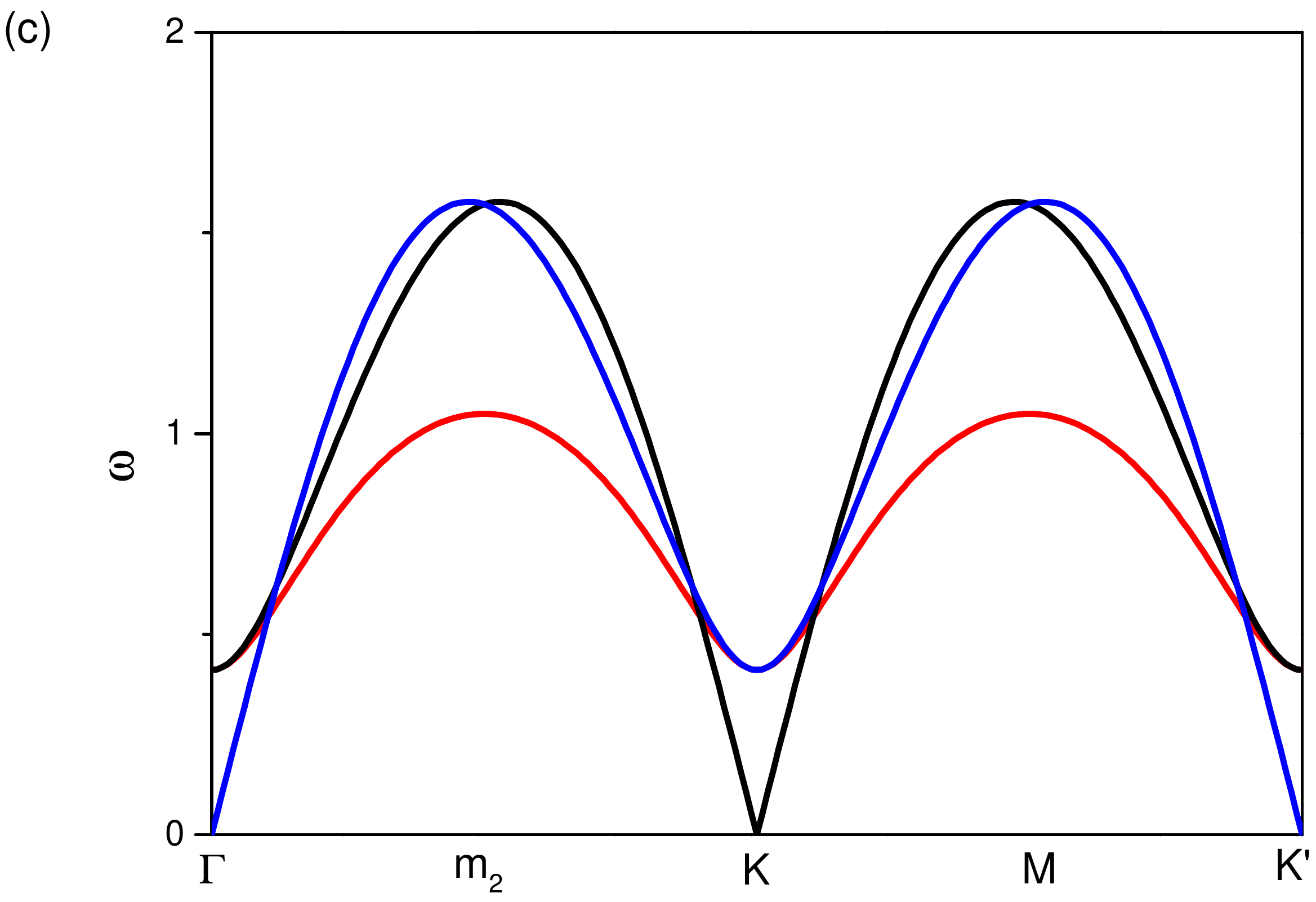}
\includegraphics[width=4.2cm,height=3cm]{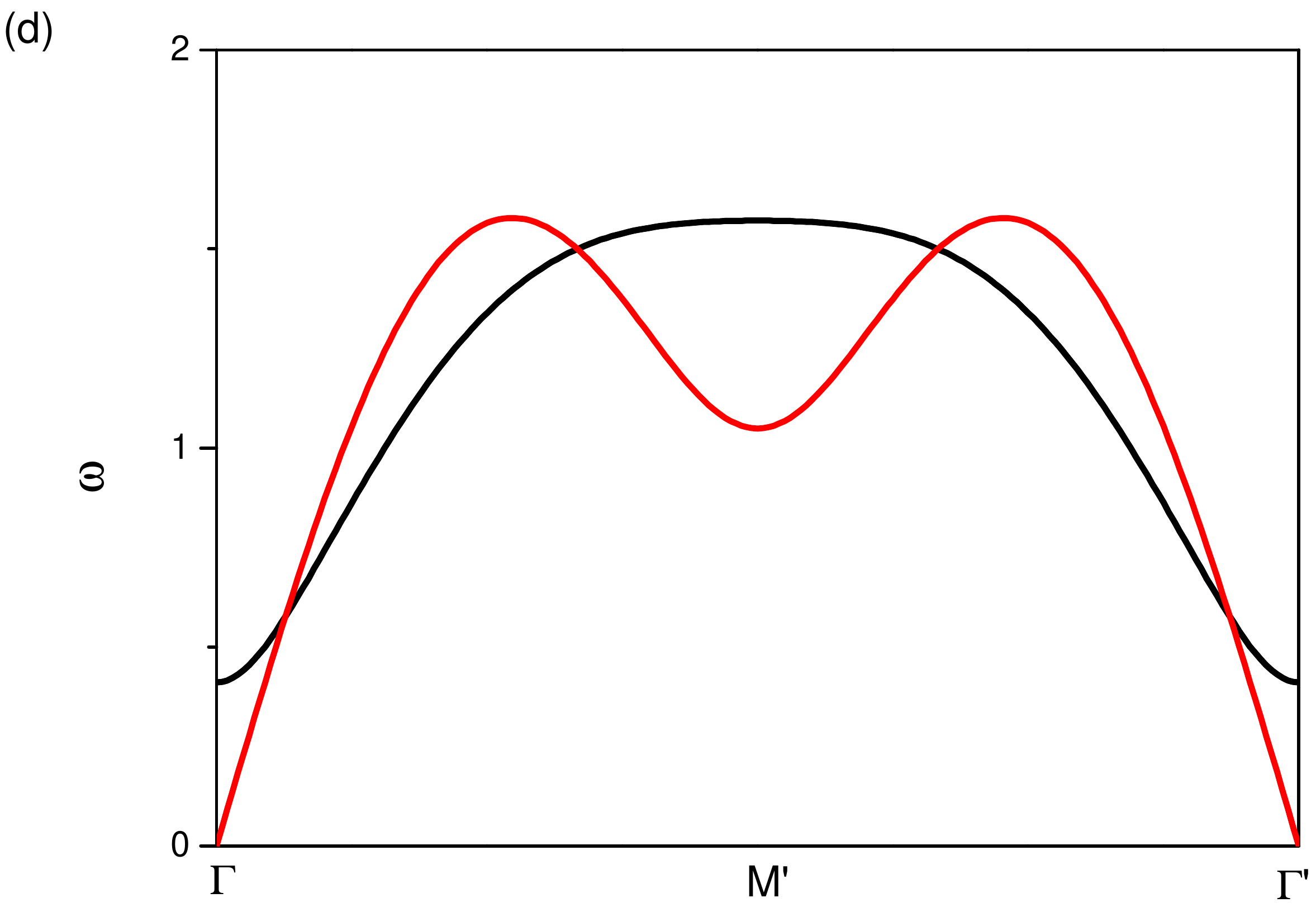}
\caption{The dispersion of the collective modes along $\Gamma-\mathbf{K}-\mathbf{K}'$(a) and $\Gamma-\mathbf{M}-\Gamma'$(b) extracted from the spin fluctuation spectrum shown in Fig.6. The dotted green lines in (a) marks the lower boundary of the spinon continuum. (c) and (d) is the magnon dispersion predicted by the LSWT. Here we assume $J^{xx}=J^{yy}$=1, $J^{zz}$=0.95.}
\end{figure}

In Fig.7, we plot the dispersion of the collective modes extracted from the spin fluctuation spectrum shown in Fig.6. Along $\Gamma-\mathbf{K}-\mathbf{K}'$(Fig.7a), we find there are indeed four collective modes in the RPA-corrected spectrum. In the vicinity of the $\Gamma$ and $\mathbf{K}$ point, the dispersion of the three modes at lower energy (shown in black, red and blue lines)show good agreement with the prediction of LSWT. However, only two of them survive around the M point, whose dispersion are strongly renormalized to exhibit roton-like minimum at the M point. The third mode first rise to touch the lower boundary of the spinon continuum(shown as dotted green line) and then vanish. Closer inspection in the mode dispersion in Fig.7a and Fig.7c indicates that there is a switching between the mode shown in the black line and that in the blue line along $\Gamma-\mathbf{K}$. This switching implies a crossing of the two modes at the M point. As we discussed earlier, this crossing is protected by the generalized translational symmetry in the 120 degree ordered phase of the spin-$\frac{1}{2}$ TLHAF.  At higher energy, there is an additional collective mode around the $\Gamma$ and the $\mathbf{K}$ point(shown in the pink line). This mode is strongly suppressed in intensity around the $\Gamma$ point as a result of intra-unit cell cancellation. The same four modes also appear in the dispersion along $\Gamma-\mathbf{M}'-\Gamma'$(Fig.7b). Here we note that the dispersion shown in black line actually corresponds to two degenerate modes. Such a mode degeneracy along $\Gamma-\mathbf{M}'-\Gamma'$ is also predicted in the LSWT. We have not plot the pink mode in Fig.7b, since its intensity is very weak.

\begin{figure}
\includegraphics[width=4.2cm,height=3cm]{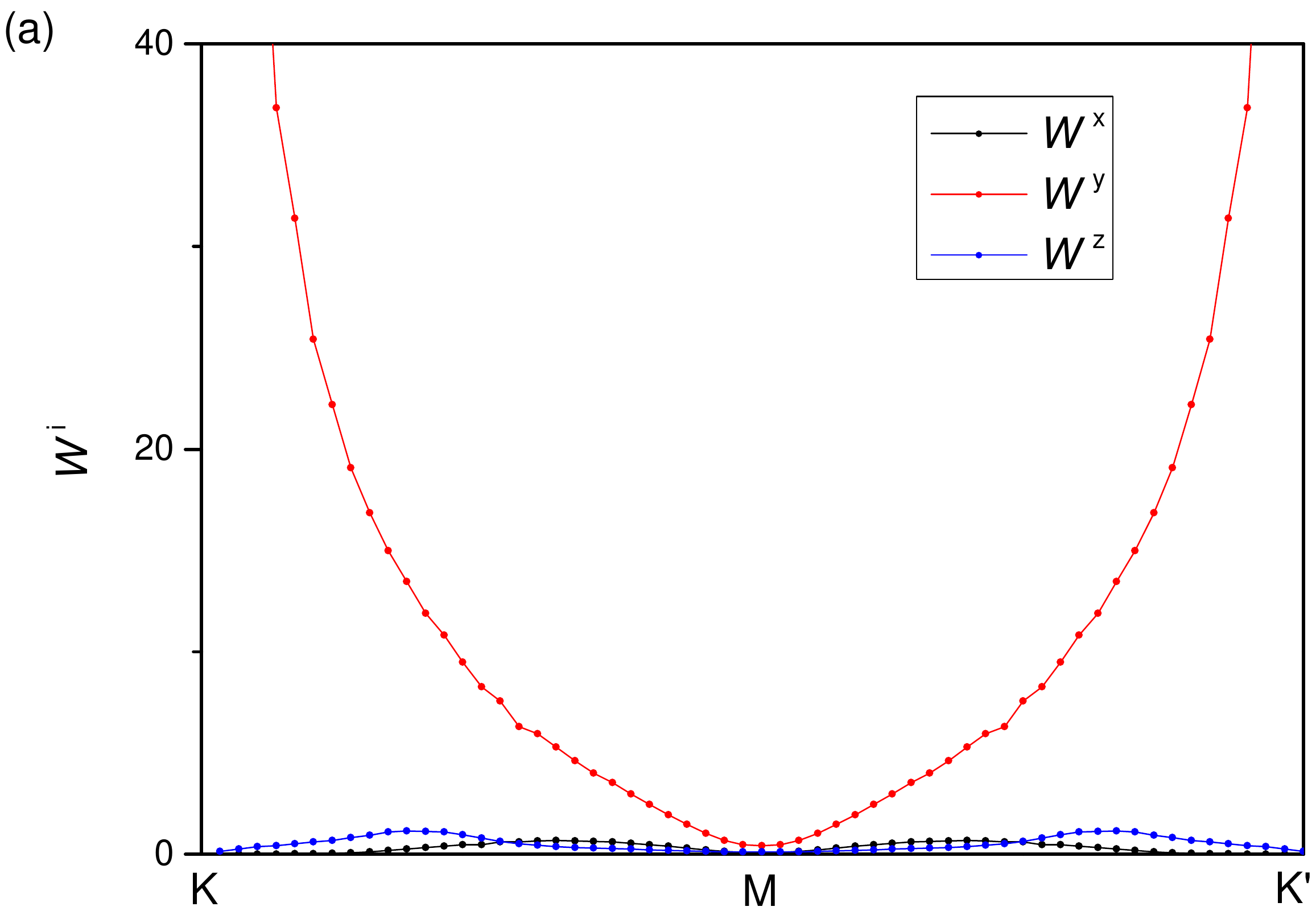}
\includegraphics[width=4.2cm,height=3cm]{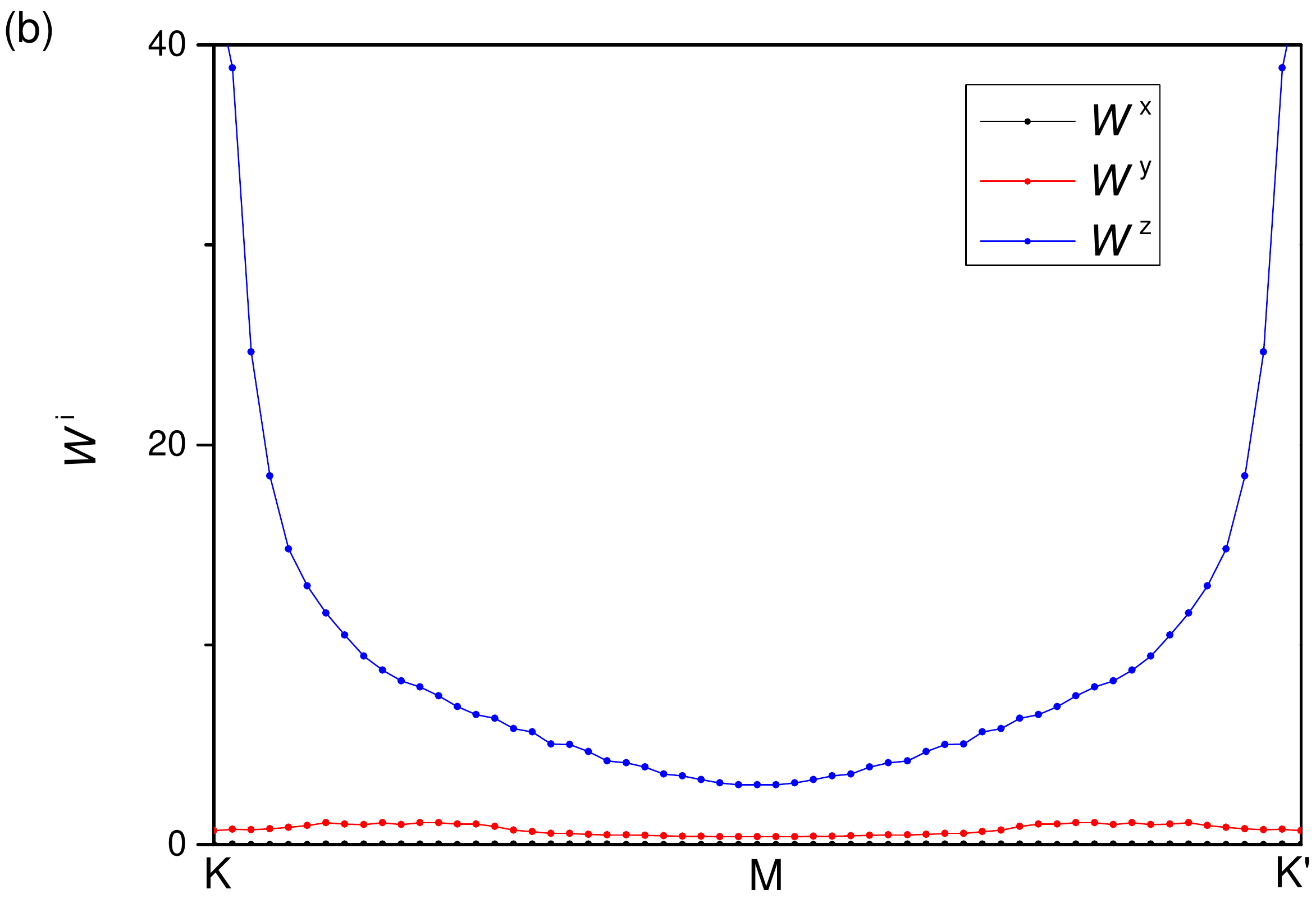}
\includegraphics[width=4.2cm,height=3cm]{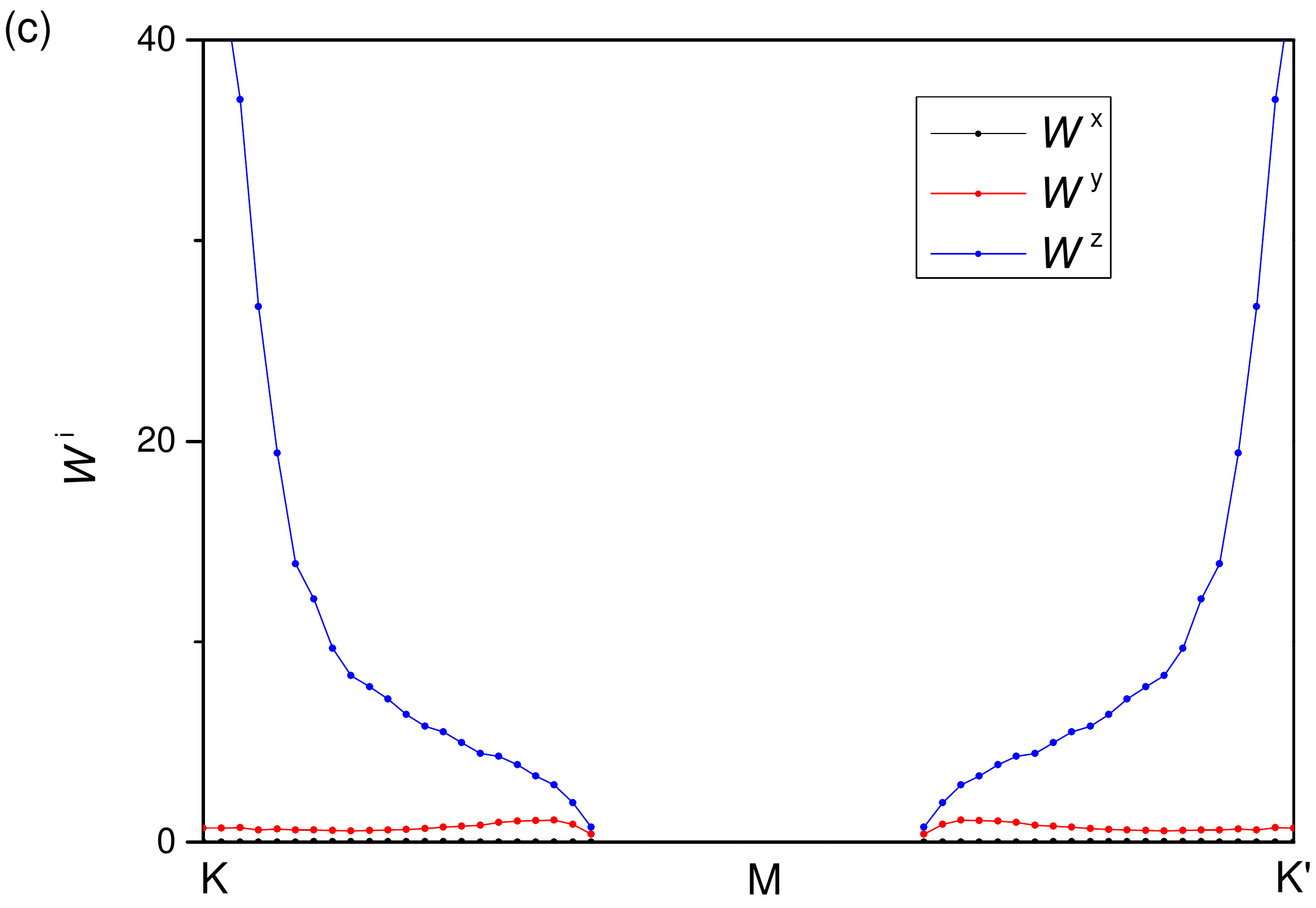}
\includegraphics[width=4.2cm,height=3cm]{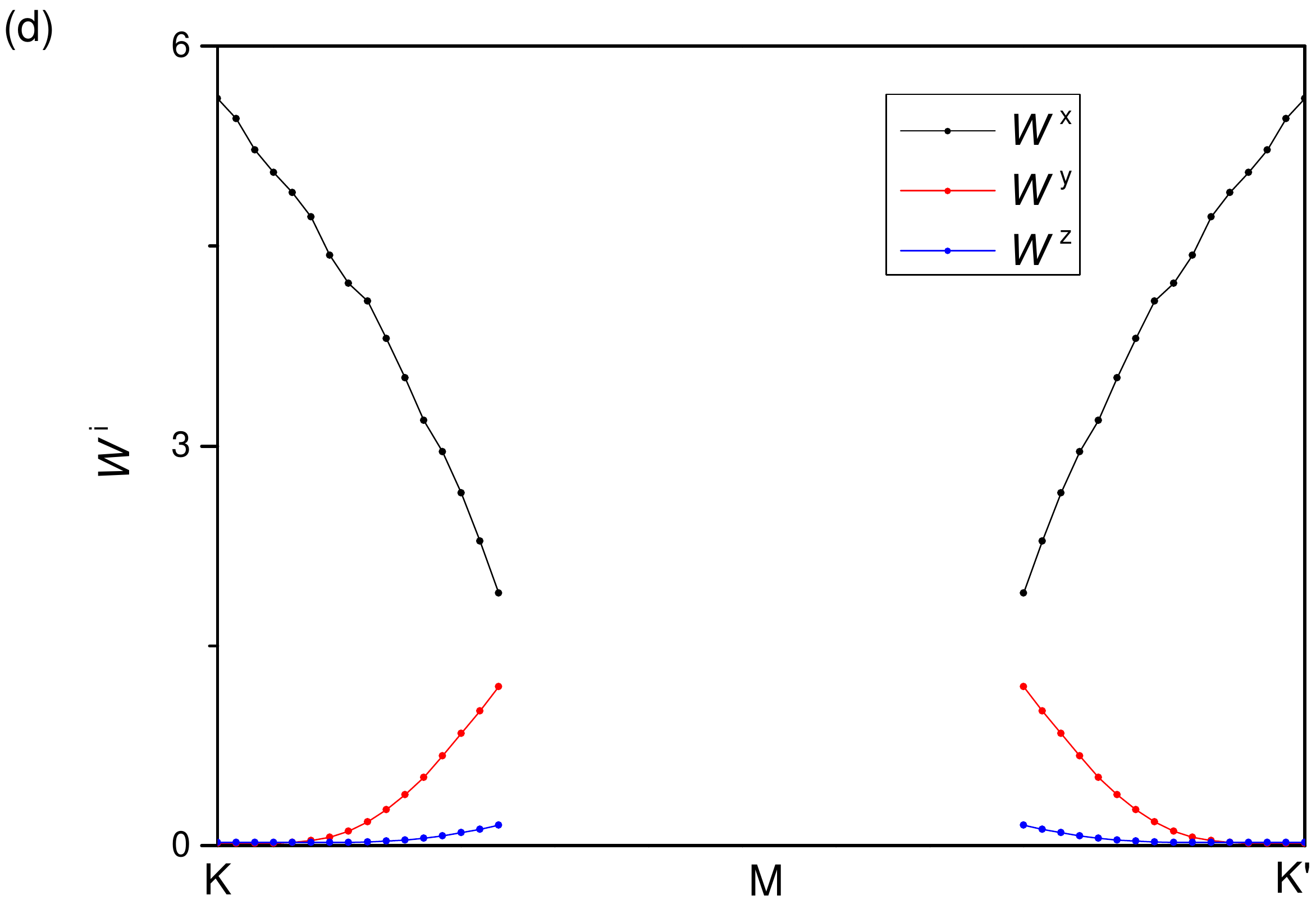}
\caption{Polarization of the four collective modes along $\mathbf{K}-\mathbf{M}-\mathbf{K}'$ shown in (a)black, (b)red, (c)blue and (d)pink curve in Fig.7a.}
\end{figure}    

\begin{figure}
\includegraphics[width=4.2cm,height=3cm]{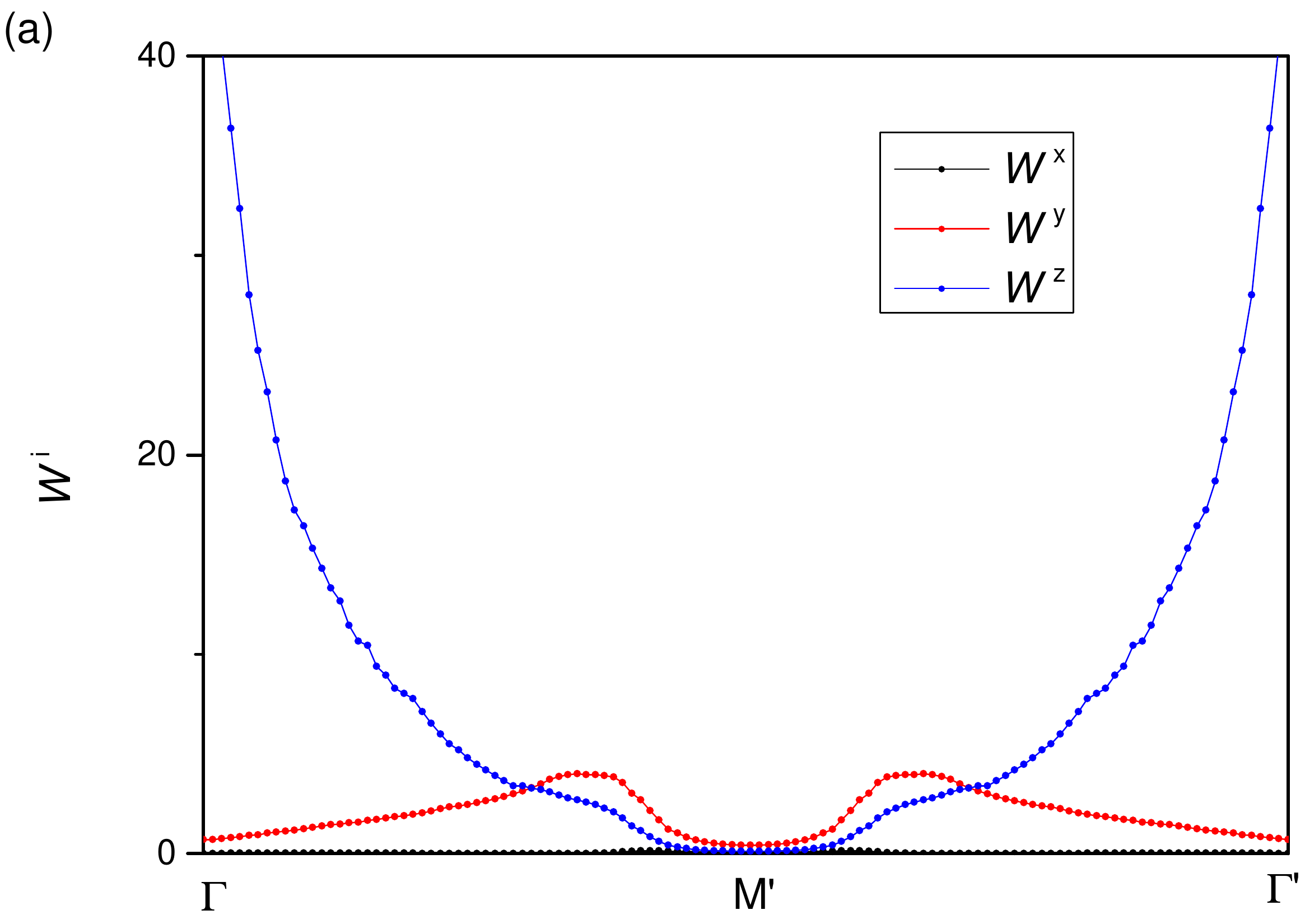}
\includegraphics[width=4.2cm,height=3cm]{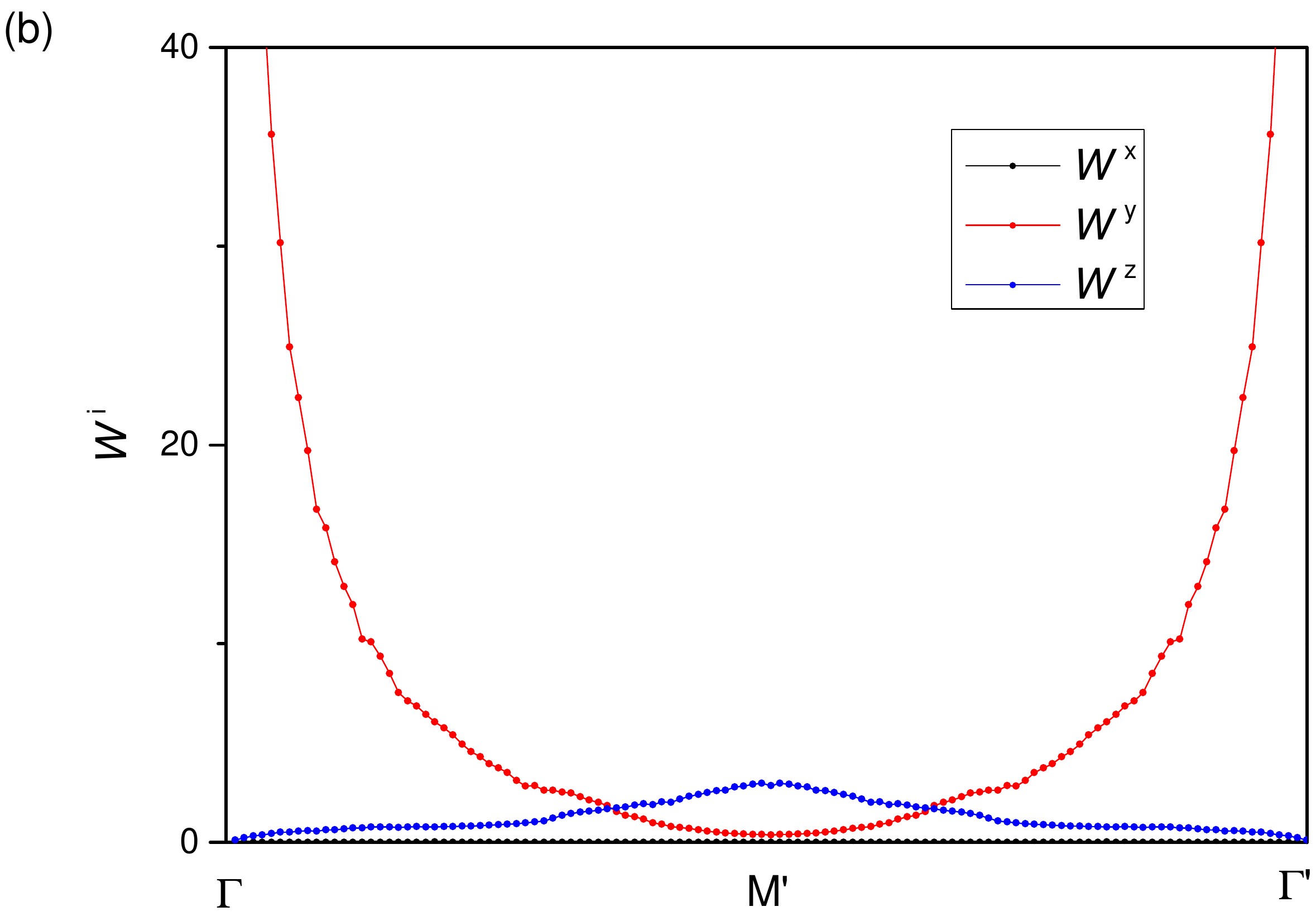}
\caption{Polarization of the two collective modes along $\Gamma-\mathbf{M}'-\Gamma'$ shown in (a)black, (b)red curve in Fig.7b.}
\end{figure}

It is clear from this analysis that the three modes at lower energy should be attributed to transverse spin fluctuation and that the fourth mode at the highest energy should be attributed to longitudinal spin fluctuation. This conclusion can be verified by studying the polarization character of the four collective modes. Such an analysis is greatly simplified by the generalized translational symmetry of the system in the 120 degree ordered phase. More specifically, spin fluctuation in different sublattices are related with each other by a 120 degree rotation in the ordering plane. We can thus focus on the polarization of spin fluctuation in a single sublattice, for example, the A sublattice. Here we note that while site A and site D belong to different sublattices in terms of spinon Hamiltonian, they belong to the same magnetic sublattice. The spectral matrix in terms of the magnetic sublattice index can be obtained from $\mathbf{R}^{i,j}_{\mu,\nu}(\mathbf{q},\omega)$ as
\begin{eqnarray}
S^{i,j}_{\tilde{\mu},\tilde{\nu}}(\mathbf{q},\omega)=\sum_{\gamma,\gamma'}e^{i(\gamma-\gamma')\mathbf{q}\cdot(\mathbf{a}_{1}+\mathbf{a}_{2})}\mathbf{R}^{i,j}_{3\gamma+\tilde{\mu},3\gamma'+\tilde{\nu}}(\mathbf{q},\omega).
\end{eqnarray}
Here $\tilde{\mu},\tilde{\nu}=1,2,3$ is the index of the three sublattices A, B and C in the magnetic unit cell, $\gamma,\gamma'=0,1$. The polarization character of the spin fluctuation mode can be read out from the spectral matrix by defining  
\begin{eqnarray}
W^{x}&=&S^{x,x}_{1,1}(\mathbf{q},\omega_{\mathbf{q}})\nonumber\\
W^{y}&=&S^{y,y}_{1,1}(\mathbf{q},\omega_{\mathbf{q}})\nonumber\\
W^{z}&=&S^{z,z}_{1,1}(\mathbf{q},\omega_{\mathbf{q}}).
\end{eqnarray}
Here $\omega_{\mathbf{q}}$ denotes the mode energy at momentum $\mathbf{q}$. More specifically, $W^{x}$ encodes the intensity of the longitudinal spin fluctuation in the given mode, $W^{y}$ and $W^{z}$ encode the intensity of the transverse spin fluctuation.

The result of  $W^{x}$, $W^{y}$ and $W^{z}$ along $\mathbf{K}-\mathbf{K}'$ and  $\Gamma-\mathbf{M}'-\Gamma'$ are presented in Fig.8  and Fig.9 respectively. From the plot one find that the mode at the highest energy is indeed dominated by longitudinal spin fluctuation. This is particularly true around the $\mathbf{K}$ and $\mathbf{K}'$ point. On the other hand, the three other modes at lower energies are all dominated by transverse spin fluctuation. Among the three transverse modes, the mode shown in the black line has dominate y-polarization around the $\mathbf{K}$ point, while the mode shown in the red and the blue lines are dominated by spin fluctuation in the z-direction around the $\mathbf{K}$ point. Around the M point, only the two lowest transverse modes survive. The spin fluctuation of these two modes have dominate z and y character.

To summarize, the spinon theory developed in this work predicts four collective modes in the MBZ. Among the four modes, there are three transverse modes and one longitudinal mode. The dispersion of the three transverse modes are strongly renormalized around the M point as a result of their coupling to the spinon continuum. In particular, the transverse mode at the highest energy has been swallowed by the spinon continuum around the M point. The two modes at lower energy are strongly softened to exhibit roton-like minimum in their dispersion. When $\mathbf{q}$ is far away from the M point, the three transverse mode recover their standard LSWT-type dispersion. The longitudinal mode is found to be the strongest around the $\mathbf{K}$ point.

Although the dispersion of the four modes predicted by the spinon theory are in close agreement with the experimental observation, the predicted distribution of spectral weight in the continuum differs significantly from that observed in the experiment. More specifically, although the third and the fourth collective mode in the RPA-corrected spectrum seem to penetrate into the continuum around the M point, there is no well defined spectral peak at the corresponding energy in the continuum at the M point. Such a failure of the theory reflects the well known limitation of the RPA approximation, according to which well-defined spectral peak can only exist outside the continuum.

\section{Conclusions and Outlooks}
In this work, we have presented a theory of the anomalous spin dynamics of the spin-$\frac{1}{2}$ TLHAF by assuming the existence of a $\pi$-flux RVB structure in its 120 degree ordered ground state. We find that with such an RVB structure and the corresponding Dirac-cone-like spinon continuum around the M point, we can explain naturally the strongly momentum dependent renormalization of the magnon dispersion observed in the INS measurement on Ba$_{3}$CoSb$_{2}$O$_{9}$. In particular, it explains naturally the origin of the roton-like minimum in the magnon dispersion around the M point. It also explains why only two magnon modes are observed in the vicinity of the M point, while the LSWT predicts three. In our theory, the magnon mode at the highest energy has been swallowed by the spinon continuum around the M point. The theory also explains naturally why the spectral continuum is the strongest around the M point. From our theory, all these spectral anomalies are interrelated and all can be attributed to the spinon transition between the two Dirac nodes.  As an additional advantage of the spinon picture, our theory predicts the existence of a strong longitudinal spin fluctuation mode around the $\mathbf{K}$ point, which seems to be indeed observed in the INS measurement.

In a quantum magnet with non-collinear ordering pattern, the polarization of the magnon mode is in general momentum dependent. This is very different from the situation in a collinear quantum magnet, in which the magnon mode is always circular polarized as a result of the conservation of the spin quantum number in the longitudinal direction. The spinon theory developed in this work can not only provide the information on the dispersion of the magnon modes, but can also provide information on their polarization character. Such information is crucial when one want to establish the correspondence between the observed spectral peaks and the theoretically predicted modes. The prediction of the longitudinal mode around the $\mathbf{K}$ point in our theory is just such an example. It is interesting to see if such a prediction can be proved by future polarized INS study. 

Despite its success in the description of the collective spin fluctuation of the system, the spinon theory presented in this work fails to describe the distribution of spectral weight in the spectral continuum. In particular, it fails to describe the two broad spectral peaks in the continuum at the M point. From the spinon theory, such broad spectral peaks in the continuum should be understood as remnant of collective spin fluctuation swallowed by the spinon continuum. However, the spectral weight predicted by the RPA theory is distributed in too broad an energy range to form well-defined spectral peak. We think this failure of the spinon theory should be attributed to the limitation of the RPA treatment adopted in this work, which is known to be inappropriate for the description of local moment behavior inside the spectral continuum. A more advanced theory is clearly needed for a thorough understanding of the spin fluctuation behavior of the spin-$\frac{1}{2}$ TLHAF. An RVB theory similar to that presented in this work, but working in the Gutzwiller projected subspace of no double occupancy, may offer such a possibility\cite{Li,Becca}.

The result presented in this work provided an example of how dramatic the presence of fractionalized spin excitation can reform the spin dynamics of a frustrated quantum magnet, even if the system host a well-defined magnetic order and that the long wavelength spin fluctuation is still well described by the LSWT. In the case of the spin-$\frac{1}{2}$ TLHAF studied in work, it is the $\pi$-flux RVB structure and the corresponding Dirac spectrum of the spinon excitation that is responsible for the spectral anomalies around the M point. The spin dynamics will be totally different if a different RVB structure, for example, an RVB structure with zero gauge flux is assumed. The anomalies in the high energy spin dynamics can thus be used as a diagnose of the nonclassical RVB correlation in the ground state of a quantum magnet. With this understanfding in mind, we expect the effort devoted in this work may shed light on mysteries of many other quantum magnets.

\begin{acknowledgments}
We acknowledge the support from the grant NSFC 11674391, the Research Funds of Renmin University of China, and the grant National Basic Research Project
2016YFA0300504. We are also grateful to Hai-Jun Liao for drawing our attention to the experimental results on Ba$_{3}$CoSb$_{2}$O$_{9}$.
\end{acknowledgments}

\end{document}